\documentclass[prd,aps,preprint,amsmath,nofootinbib,amssymb,eqsecnum,showkeys,tightenlines]{revtex4-1}

\pdfoutput=1

\usepackage{graphicx}
\usepackage{xcolor}

\begin{document}
\title{Neutrino lines and photon continua from cascade dark matter decay}

\author{Jun Guo}
\email{jguo_dm@jxnu.edu.cn}
\affiliation{College of Physics, Jiangxi Normal University,
Nanchang 330022, China}
\affiliation{National Gravitation Laboratory,
Huazhong University of Science and Technology,
Wuhan 430074, China}

\author{Shi-ying Zhao}
\affiliation{College of Physics, Jiangxi Normal University,
Nanchang 330022, China}

\author{Bin Zhu}
\email{zhubin@mail.nankai.edu.cn}
\affiliation{Department of Physics, Yantai University,
Yantai 264005, China}

\begin{abstract}
We investigate the two-body decay of fermionic dark matter,
$\chi(\bar{\chi})\to X+\nu(\bar{\nu})$, where the light mediator
$X$ subsequently decays into photons. We consider two benchmark
models: an axion-like particle with $a\to\gamma\gamma$, and a
kinetically mixed dark vector with $A'\to3\gamma$. This decay
topology produces a monochromatic neutrino line from the primary
decay together with a broad secondary photon continuum. A key
feature of the scenario is that the photon signal depends on the
mediator decay length, whereas the neutrino line is produced
promptly and is insensitive to the subsequent propagation of $X$.
We derive dark matter lifetime limits from current and projected
MeV gamma-ray and neutrino searches, including both Galactic and
delayed extragalactic photon contributions. We find that photon
constraints generally dominate for short-lived mediators, while
neutrino-line searches can become competitive or provide the
leading sensitivity in regions where mediator propagation
substantially suppresses the photon signal. This conclusion remains
stable under conservative extragalactic-only limits and a simplified
treatment of Galactic spatial smearing.
\end{abstract}

\maketitle
\section{Introduction}
Dark-matter searches have historically emphasized weak-scale candidates~\cite{Bertone:2004pz}. The absence of a confirmed signal in this regime, together with broad theoretical interest in light dark sectors, has motivated increasing attention to sub-GeV dark matter~\cite{Boehm:2003hm,Boehm:2003ha,Essig:2013lka,Yi:2026fmf,Guo:2024sqh,Guo:2023kqt,Liang:2024xcx,Yin:2024lla,Ruz:2024gkl,Roy:2023omw,Langhoff:2022bij,Iles:2024zka,Du:2024afd,Chen:2024ekh,Guo:2025dlk,Sun:2025gyj,Lu:2023cet,Guo:2023hyp,Su:2023zgr}. In this mass range, decaying dark matter (DDM) provides a well-motivated framework for indirect searches~\cite{Akita:2022lit}. If dark-matter stability is only approximate, small symmetry-breaking effects or higher-dimensional operators can naturally induce lifetimes much longer than the age of the Universe~\cite{Ibarra:2013cra}. Despite such long lifetimes, DDM can still generate observable indirect-detection signals~\cite{Palomares-Ruiz:2007egs}, with the resulting signal flux scaling linearly with the dark-matter density.

In many realistic dark-sector models, DM does not decay directly into Standard Model particles but proceeds through an intermediate state. Such multi-step processes can qualitatively alter the resulting observable signatures, especially when the intermediate particle is long-lived, leading to novel multi-messenger phenomenology. A generic possibility is the two-step decay
\begin{equation}
\chi \to X + \nu\,\,(\bar{\chi}\to X + \bar{\nu}), \qquad X \to {\rm SM},
\label{eq:process}
\end{equation}
where $X$ is a light mediator such as an axion-like particle (ALP)~\cite{Caputo:2022dkz,Adams:2022pbo,Janish:2023kvi} or a dark vector~\cite{Fabbrichesi:2020wbt,Caputo:2021eaa, Linden:2024fby,Rizzo:2018ntg}. This topology combines a metastable parent coupled to the neutrino sector ~\cite{Falkowski:2009yz,Coy:2020wxp} with a long-lived daughter whose propagation can modify the electromagnetic signal ~\cite{Chu:2017vao}. The two lifetimes need not have the same microscopic origin: the parent may be stabilized by an approximate symmetry or a high-scale portal, whereas the daughter lifetime can be controlled by a weak visible-sector coupling or by kinematic closure of faster decay channels.

Indirect searches for DDM typically rely on SM product signatures, particularly gamma rays, which offer excellent spectral and spatial information. In the MeV energy range, instruments such as COMPTEL~\cite{Strong:1998ck, Essig:2013goa} and INTEGRAL~\cite{ Siegert:2022jii, Berteaud:2022tws} have provided foundational constraints, while future missions like AMEGO-X aim to significantly enhance sensitivity. Nevertheless, this energy range—often referred to as the "MeV gap"~\cite{ODonnell:2024aaw, Saha:2025wgg}—remains challenging due to formidable astrophysical backgrounds and detector performance limitations. We focus on the regime in which $X$ propagates over Galactic or cosmological distances before decaying. In this regime, the photons are produced in the secondary decay of
$X$ and therefore depend on its propagation. For the Galactic component we use a collinear benchmark, supplemented by a delayed extragalactic (EG) contribution, and test the resulting photon/neutrino classification with EG-only limits and a spatial-smearing calculation.

Neutrino-line signatures from decaying DM have been studied in Majoron, seesaw, and neutrino-portal scenarios ~\cite{Garcia-Cely:2017oco,Coy:2020wxp,Patel:2019zky}, and correlated neutrino and electromagnetic signals have also been considered ~\cite{ElAisati:2015qec,Arguelles:2022nbl,Das:2024tmx}. Since the primary neutrino is produced promptly, we study how the relative sensitivity of the two messengers changes when the electromagnetic signal is produced by a propagating daughter. For the benchmarks $a\to\gamma\gamma$ and $A'\to3\gamma$, we identify the neutrino-dominated regions and examine their dependence on the mediator lifetime, the mass hierarchy $m_\chi/m_X$, and the visible coupling.

The paper is organized as follows: In Section~\ref{sec:model}, we introduce the DDM models and study the characteristic photon and neutrino spectra. In Section~\ref{sec:photon}, we derive constraints from gamma-ray observations, followed by the neutrino signal analysis in Section~\ref{sec:neutrino}. The multi-messenger synergy and the identification of neutrino-dominant parameter space are discussed in Section~\ref{sec:synergy}. Finally, we present our results and conclusions in Section~\ref{sec:conclusion}.

\section{The simplified decay DM model}
\label{sec:model}
We adopt a simplified low-energy description~\cite{Dev:2025tdv} and supplement it with illustrative neutrino-portal constructions that demonstrate how the required effective vertices and the hierarchy between the parent and mediator lifetimes can arise. These constructions are not imposed as complete ultraviolet models in the numerical analysis; instead, the primary decay is parameterized by $\tau_\chi$, while the mediator lifetime is controlled by its visible coupling in the benchmark scenarios below.

We first summarize the kinematics of the process in Eq.~\ref{eq:process}. The energy of the monochromatic neutrino is determined by:
\begin{equation}
E_{\nu0} = \frac{m_\chi}{2} \left( 1 - \frac{1}{k^2} \right) \, ,
\label{eq:E_nu}
\end{equation}
where we have introduced the mass ratio $k \equiv m_\chi / m_X$. This parameter $k$ plays a crucial role in our analysis: for $k \gg 1$, the neutrino carries approximately half of the DM mass, while for $k \to 1$, the decay becomes non-relativistic, producing a soft neutrino and a nearly stationary mediator. The corresponding monochromatic line spectrum is 
\begin{equation}
\frac{dN_\nu}{dE_\nu} = \delta(E_\nu - E_{\nu0})\,,
\label{eq:dndenu}
\end{equation}
The secondary photon signal depends on the propagation of the mediator. Its Lorentz boost factor is
\begin{equation}
\gamma_X = \frac{E_{X,0}}{m_X} = \frac{1}{2} \left( k + \frac{1}{k} \right) \, .
\label{eq:gamma_X}
\end{equation}
The characteristic decay length in the laboratory frame is $\lambda_X = \beta_X \gamma_X \tau_X $. In the Galactic-collinear benchmark, a mediator produced at a line-of-sight distance $s$ contributes only if it decays before reaching the observer. The corresponding decay probability is~\cite{Agashe:2022phd}
\begin{equation}
P_{\rm col}(\Gamma_X,s)=1-\exp\left(-\frac{s}{\lambda_X}\right) .
\label{eq:decay_prob_collinear}
\end{equation}

This propagation factor suppresses the Galactic photon signal when the mediator decay length becomes comparable to or larger than Galactic scales. We consider two representative visible decay channels for $X$:
\begin{itemize}
    \item \textbf{Axion-like particle (ALP) mediator ($X=a$):} The ALP decays into two photons ($a \to \gamma\gamma$) via the dimension-5 operator $\frac{1}{4}g_{a\gamma\gamma} a F_{\mu\nu} \tilde{F}^{\mu\nu}$.
    
    \item \textbf{Kinetically mixed dark-vector mediator ($X=A'$):} We focus on $m_{A'}<2m_e$, for which the tree-level $A'\to e^+e^-$ channel is kinematically forbidden. The leading visible decay is then the one-loop process $A'\to3\gamma$. The combination of the kinetic-mixing suppression, the loop-induced decay, and the steep low-mass scaling of the three-photon width can naturally lead to macroscopic or astrophysical decay lengths~\cite{McDermott:2017qcg}.
\end{itemize}

\subsection{Axion-like particle mediator}
\label{subsec:axion}
The Lorentz structure of the primary transition depends on the spin of the mediator. For the ALP benchmark, we use the following low-energy effective interaction, valid below the electroweak scale,
\begin{equation}
\mathcal L_{\chi\nu a}=a\bar\chi\left(g^L_{\chi\nu a}P_L+g^R_{\chi\nu a}P_R\right)\nu+\mathrm{h.c.}
\label{eq:L_chi_nu_a}
\end{equation}
In the phenomenological analysis only the total primary lifetime $\tau_\chi$ enters, so the detailed chiral structure of the transition is not important. For notational simplicity, one may keep a single effective coupling $g_{\chi\nu a}$.

To illustrate that the effective interaction in Eq.~\eqref{eq:L_chi_nu_a} can arise in a conventional neutrino-portal setting, consider a dark sector with an approximate global symmetry $U(1)_A$ that is spontaneously broken. A minimal set of fields relevant for generating the portal interaction includes a complex scalar
\begin{equation}
\Phi=\frac{1}{\sqrt{2}}(f_a+\rho)\exp\left(\frac{i a}{f_a}\right),
\end{equation}
whose phase is the ALP $a$. We assume that $U(1)_A$ is approximate, with a small explicit breaking generating a nonzero ALP mass $m_a$ without qualitatively modifying the portal structure relevant here. We further introduce a sterile neutral fermion $N_R$ and a vector-like dark fermion $\Psi$. With suitable charge assignments under the broken $U(1)_A$, the relevant interactions may be written schematically as
\begin{equation}
\mathcal L\supset-y_\nu\,\bar L\tilde H N_R-y_D \Phi\,\bar\Psi_L N_R+\mathrm{h.c.},
\label{eq:alp_uv_schematic}
\end{equation}
where $L$ and $H$ are the SM lepton and Higgs doublets. After electroweak and $U(1)_A$ symmetry breaking, these terms mix the active neutrino, the sterile state, and the dark fermion. The DDM particle $\chi$ may then be identified with a heavy mass eigenstate dominantly composed of $\Psi$.

Expanding the phase of $\Phi$ gives
\begin{equation}
y_D\Phi\,\bar\Psi_L N_R\supset m_D'\bar\Psi_L N_R+i\frac{m_D'}{f_a}a\bar\Psi_L N_R,
\qquad
m_D'=\frac{y_D f_a}{\sqrt{2}} .
\end{equation}
After rotating to the mass-eigenstate basis, this interaction induces off-diagonal ALP couplings between the heavy dark fermion and the light neutrino. Parametrically,
\begin{equation}
g_{\chi\nu a}^{L,R}\sim\frac{m_D'}{f_a}\,U_{\chi\Psi}^{\ast}U_{\nu N},
\end{equation}
up to chiral mixing factors and order-one phase conventions. Thus, the decay $\chi\to a\nu$ can naturally arise from neutrino-portal mixing. In the following analysis we treat the primary lifetime $\tau_\chi$ as a phenomenological parameter.

The visible decay $a\to\gamma\gamma$ can be generated independently by heavy vector-like charged fermions carrying the broken $U(1)_A$ charge. After integrating them out, one obtains the standard anomalous interaction
\begin{equation}
\mathcal L_{a\gamma\gamma}=\frac{1}{4}g_{a\gamma\gamma}a F_{\mu\nu}\tilde F^{\mu\nu},
\end{equation}
which controls the ALP decay length relevant for the photon signal. The corresponding visible decay width is~\cite{Cadamuro:2011fd, Saha:2025any}
\begin{equation}
\Gamma_{a\to\gamma\gamma}=\frac{g_{a\gamma\gamma}^2 m_a^3}{64\pi}.
\end{equation}
For an ALP interpreted as a pseudo-Nambu--Goldstone boson, the small photon coupling also admits a symmetry-based interpretation. Schematically, the anomalous coupling can be written as $g_{a\gamma\gamma}\sim\frac{\alpha}{2\pi}\frac{C_\gamma}{f_a}$, where $C_\gamma$ is model dependent. A large $f_a$ therefore provides a simple origin for a long-lived ALP, independently of the suppression responsible for the primary DM decay. In the numerical analysis we nevertheless treat $g_{a\gamma\gamma}$ as an independent effective coupling.

The neutrino-portal sector can also induce invisible ALP decays. We restrict to mass orderings that close channels involving the heavier neutral fermions and to the region 
$\Gamma(a\to{\rm invisible})\ll\Gamma(a\to\gamma\gamma)$. 
Accordingly, ${\rm Br}(a\to\gamma\gamma)\simeq1$ in our benchmark.

In the axion rest frame, the decay $a \to \gamma\gamma$ produces two monochromatic photons,
\begin{equation}
\frac{d N_\gamma}{d E'_\gamma}= 2\delta\left(E'_\gamma-\frac{m_a}{2}\right).
\end{equation}
Boosting to the DM rest frame, the photon spectrum from $a \to \gamma\gamma$ is therefore a broad continuum whose shape is fully determined by
\begin{equation}
\frac{d N_\gamma}{d E_\gamma}= \frac{2}{E_+-E_-}\Theta(E_\gamma-E_-)\Theta(E_+-E_\gamma),
\end{equation}
with the endpoints
\begin{equation}
E_\pm = \frac{E_a}{2}(1\pm\beta_a), \qquad \beta_a = \sqrt{1-\frac{m_a^2}{E_a^2}}.
\end{equation}
The boosted spectrum is a box of width $\Delta E=m_\chi(k^2-1)/(2k^2)$. It approaches a narrow line as $k\to1$ and spans approximately $0<E_\gamma<m_\chi/2$ for $k\gg1$, as shown in Fig.~\ref{fig:mediator_spectrum}.

\subsection{Dark-vector mediator benchmark}
\label{subsec:darkphoton}
We also consider the case where the mediator is the vector boson, with the simplified model given as
\begin{equation}
\mathcal L_{\chi\nu A'}=g_{\chi\nu A'}\,A'_\mu\,\bar\chi\gamma^\mu P_L\nu+\mathrm{h.c.}.
\end{equation}
The kinetic mixing is not responsible for the primary $\chi\to A'\nu$ transition, since a kinetically mixed vector does not directly couple to SM neutrinos. We instead regard this vertex as an effective neutrino-portal interaction. To illustrate how such a coupling can arise, consider a broken dark gauge symmetry $U(1)_D$ with a dark Higgs field $S$, a vector-like dark fermion $\Psi$ charged under $U(1)_D$, and a sterile neutral fermion $N_R$. With suitable charge assignments, the relevant interactions can be written schematically as
\begin{equation}
\mathcal L \supset-\frac{\varepsilon}{2}F'_{\mu\nu}F^{\mu\nu}-y_\nu \bar L\tilde H N_R-y_D S\bar\Psi_L N_R+{\rm h.c.},
\label{eq:uv_dark_vector_schematic}
\end{equation}
together with the dark gauge interaction
\begin{equation}
\mathcal L \supset g_D A'_\mu \bar\Psi\gamma^\mu\Psi .
\end{equation}
After electroweak and $U(1)_D$ symmetry breaking, the active neutrino, the sterile state, and the dark fermion mix. 
The DDM particle $\chi$ can then be identified with a heavy mass eigenstate dominantly composed of $\Psi$. 
Rotating the dark gauge current to the mass-eigenstate basis generates the effective transition
\begin{equation}
\mathcal L_{\chi\nu A'} = \sum_i g_{\chi\nu_i A'}A'_\mu\bar\chi\gamma^\mu P_L\nu_i + {\rm h.c.},
\label{eq:effective_chi_nu_Ap}
\end{equation}
with the parametric scaling
\begin{equation}
g_{\chi\nu_i A'}
\sim
g_D\,U_{\chi\Psi}^{\ast}U_{\nu_i\Psi}.
\end{equation}

Thus, the primary decay $\chi\to A'\nu_i$ arises from neutrino-portal mixing, while the kinetic-mixing parameter $\varepsilon$ controls the visible interactions and lifetime of $A'$. An important feature of this realization is that the primary transition and the invisible decay of the daughter carry different powers of the small active--dark mixing. For a dark-matter state dominantly composed of $\Psi$, $|U_{\chi\Psi}|\simeq1$, one has parametrically
\begin{equation}
g_{\chi\nu_i A'} \sim g_D U_{\nu_i\Psi}, \qquad g_{\nu_i\nu_j A'} \sim g_D U_{\nu_i\Psi}^{*}U_{\nu_j\Psi}.
\end{equation}
Thus the primary width scales as $|U_{\nu\Psi}|^2$, whereas $A'\to\nu\bar\nu$ is suppressed by $|U_{\nu\Psi}|^4$. We therefore treat $\tau_\chi$ as an independent phenomenological parameter and restrict to the visible-decay-dominated region $\Gamma(A'\to\nu\bar\nu)\ll\Gamma(A'\to3\gamma)$. Channels containing heavier neutral fermions are assumed to be kinematically closed, so that ${\rm Br}(A'\to3\gamma)\simeq1$.

The dark-vector decays into three photons with the loop-induced width, analogous to the kinetically mixed dark-photon case~\cite{Pospelov:2008jk}
\begin{align}
\Gamma_{A'\to 3\gamma} &= \Gamma_{\rm EH}\times f_{\rm{loop}}(m_{A'}) \nonumber\\
&\approx \frac{17 \varepsilon^2 \alpha_{\rm EM}^4}{11664000\, \pi^3} \frac{ m_{A'}^9}{m_e^8}\left[1+\sum_{n=1}^\infty c_n \left(\frac{m_{A'}^2}{m_e^2}\right)^n\right],
\label{eq:Gamma_Adp}
\end{align}
We write the result as the Euler--Heisenberg width $\Gamma_{\rm EH}$ multiplied by an enhancement factor $f_{\rm loop}(m_{A'})$ that depends on the dark-vector mass; the coefficients $c_n$ are given in Table I of ~\cite{McDermott:2017qcg}. The sub-$2m_e$ region therefore provides a particularly motivated long-lived window. Once $A'\to e^+e^-$ is kinematically closed, the leading visible width is loop induced and scales as $\Gamma_{A'\to3\gamma}\propto\varepsilon^2m_{A'}^9$ in the Euler--Heisenberg limit. The full loop result becomes important near the $e^+e^-$ threshold~\cite{McDermott:2017qcg}. Thus, astrophysical decay lengths arise over an appreciable region of parameter space without introducing a separate small decay coupling.

In the rest frame of the dark-vector, the photon spectrum from $A'\to 3\gamma$ can be written in terms of $x=2E'_\gamma/m_{A'}$ as~\cite{Pospelov:2008jk}
\begin{equation}
\frac{dN_\gamma}{dx}=\frac{x^3}{17}\left(1715-3105x+\frac{2919}{2}x^2\right),
\label{eq:distHE}
\end{equation}
This spectrum vanishes in the soft-photon limit $x\to0$, has support $0\le x\le1$, and integrates to three photons. The shape is therefore neither monochromatic nor box-like even before the boost to the dark-matter rest frame.

The boosted continuum used in our numerical pipeline is then
\begin{equation}
\frac{d N_\gamma}{d E_\gamma}
= \int_{E'_{\rm min}}^{E'_{\rm max}}
\frac{d E'_\gamma}{2\gamma_{A'}\beta_{A'}E'_\gamma}
\frac{d N_\gamma}{d E'_\gamma},
\end{equation}
with
\begin{equation}
E'_{\rm min} = \frac{E_\gamma}{\gamma_{A'}(1+\beta_{A'})},
\qquad
E'_{\rm max} = \min\!\left[\frac{m_{A'}}{2},\frac{E_\gamma}{\gamma_{A'}(1-\beta_{A'})}\right],
\quad
\frac{d N_\gamma}{d E'_\gamma} = \frac{2}{m_{A'}}\frac{dN_\gamma}{dx}.
\end{equation}
\begin{figure}[htbp]
\centering
\includegraphics[width=0.49\textwidth]{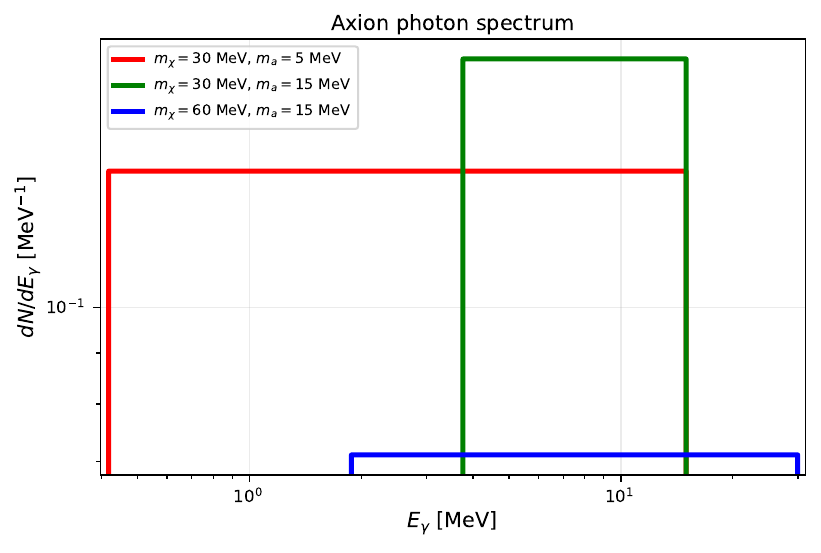}
\includegraphics[width=0.49\textwidth]{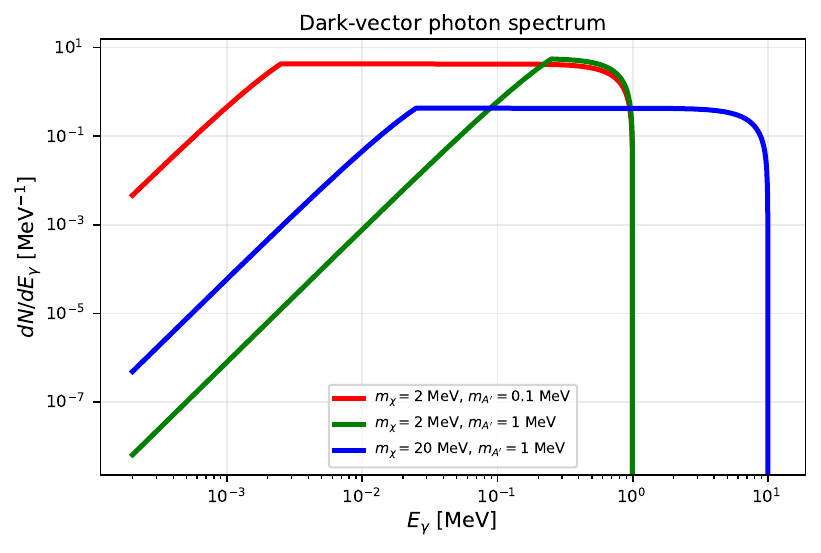}
\caption{Differential photon spectra $dN/dE$ for the axion-like particle (left) and dark-vector (right) mediator scenarios. Left panel: The ALP-mediated spectrum exhibits a characteristic box-like shape, where the width $\Delta E$ is determined by the boost factor $\gamma_a$. For smaller $k$, the box narrows and approaches a monochromatic line. Right panel: The dark-vector case produces a smoother continuum due to the three-body decay $A' \to 3\gamma$.}
\label{fig:mediator_spectrum}
\end{figure}
For $m_{A'}\ll m_\chi$, the dark vector is highly boosted and the three-photon spectrum extends up to $E_\gamma\sim m_\chi/2$. Unlike the ALP box, however, the boosted spectrum remains smooth because of the three-body rest-frame distribution, as illustrated in Fig.~\ref{fig:mediator_spectrum}.

Before turning to the indirect-detection constraints, we briefly clarify the cosmological scope of our analysis. We do not include constraints associated with an early-Universe population of the mediator $X$. In a standard thermal history, the same visible-sector interactions considered here can in principle produce ALPs or dark vectors in the primordial plasma, and sufficiently long-lived populations may be constrained by BBN, CMB observables, or late electromagnetic energy injection~\cite{Balazs:2022tjl,Caputo:2025avc,Xu:2025wlq}. The importance of these constraints depends on the mediator production history and, more generally, on the thermal history and ultraviolet realization of the dark sector.

\section{Constraints from gamma-ray observations}
\label{sec:photon}

The Galactic photon flux in the collinear approximation is
\begin{equation}
\frac{d\Phi_{\gamma}^{\rm{Gal,col}}}{dE_0} = \frac{1}{4\pi m_\chi\tau_\chi} \frac{dN_\gamma}{dE_0} \times D_{\rm eff},
\label{eq:photon_flux}
\end{equation}
where the mediator is assumed to propagate along the observed line of sight. We use a factorized line-of-sight benchmark for the Galactic component. The mediator decay probability is evaluated by assuming that $X$ propagates along the parent line of sight toward the observer, while the secondary-photon spectrum is retained as the angle-averaged boosted spectrum. We therefore neglect the angular redistribution of the decay photons and the associated angle--energy correlations. The propagation-induced suppression is encapsulated in the effective $D$-factor, which is defined as:
\begin{equation}
D_{\rm eff} = \int_{\rm ROI} d\Omega \int_0^{s_{\rm max}} ds \, \rho(r)  P_{\rm col}(\Gamma_X,s),
\label{eq:deff}
\end{equation}
Here $s$ is the line-of-sight distance. In the prompt limit $D_{\rm eff}$ approaches the usual decay $D$-factor, whereas for $\lambda_X\gg s$, $P_{\rm col}\simeq s/\lambda_X\propto\Gamma_X$, suppressing the Galactic photon signal. The Galactic-collinear calculation factorizes the decay probability from the angle-averaged photon spectrum and should not be interpreted as a full angular--energy transport treatment. We therefore use it as a benchmark Galactic component and test the dominance classification independently with the EG-only limit and the spatial-smearing normalization check in Appendices~\ref{app:EG} and~\ref{app:spatial_smearing}.

For the DM density profile $\rho(r)$, which denotes the total density of $\chi$ and $\bar{\chi}$, $r = \sqrt{s^2 + R^2 - 2sR \cos{l}\cos{b}}$ is the radial distance from the Galactic Center, and $R=8.12\ \rm{kpc}$ is the distance from the Sun to the Galactic Center, $(l, b)$ is the Galactic coordinates. We adopt a generalized NFW profile~\cite{He:2019svf},
\begin{equation}
    \rho_{\rm gNFW} = \frac{\rho_s}{(r/r_s)^\gamma(1+r/r_s)^{3-\gamma}}
    \label{eq:gnfw}
\end{equation}
which provides a better description of cosmological hydrodynamic simulations that include baryons. We take $\gamma=1.2$~\cite{Fermi-LAT:2017opo} and $r_s=20~{\rm kpc}$. The normalization $\rho_s=0.25~{\rm GeV\,cm^{-3}}$ is chosen to reproduce the local DM density $\rho(R)=0.4~{\rm GeV\,cm^{-3}}$ at $R=8.12~{\rm kpc}$.

For comparison with diffuse gamma-ray data, we also define the ROI-averaged intensity rather than the ROI-integrated flux, given by
\begin{equation}
    \frac{d^2\Phi_{\gamma}^{\rm{Gal,col}}}{dE_0\,d\Omega} =\frac{1}{\Delta\Omega}\frac{d\Phi_{\gamma}^{\rm{Gal,col}}}{dE_0}
\end{equation}

\begin{figure}[htbp]
\centering
\includegraphics[width=0.49\textwidth]{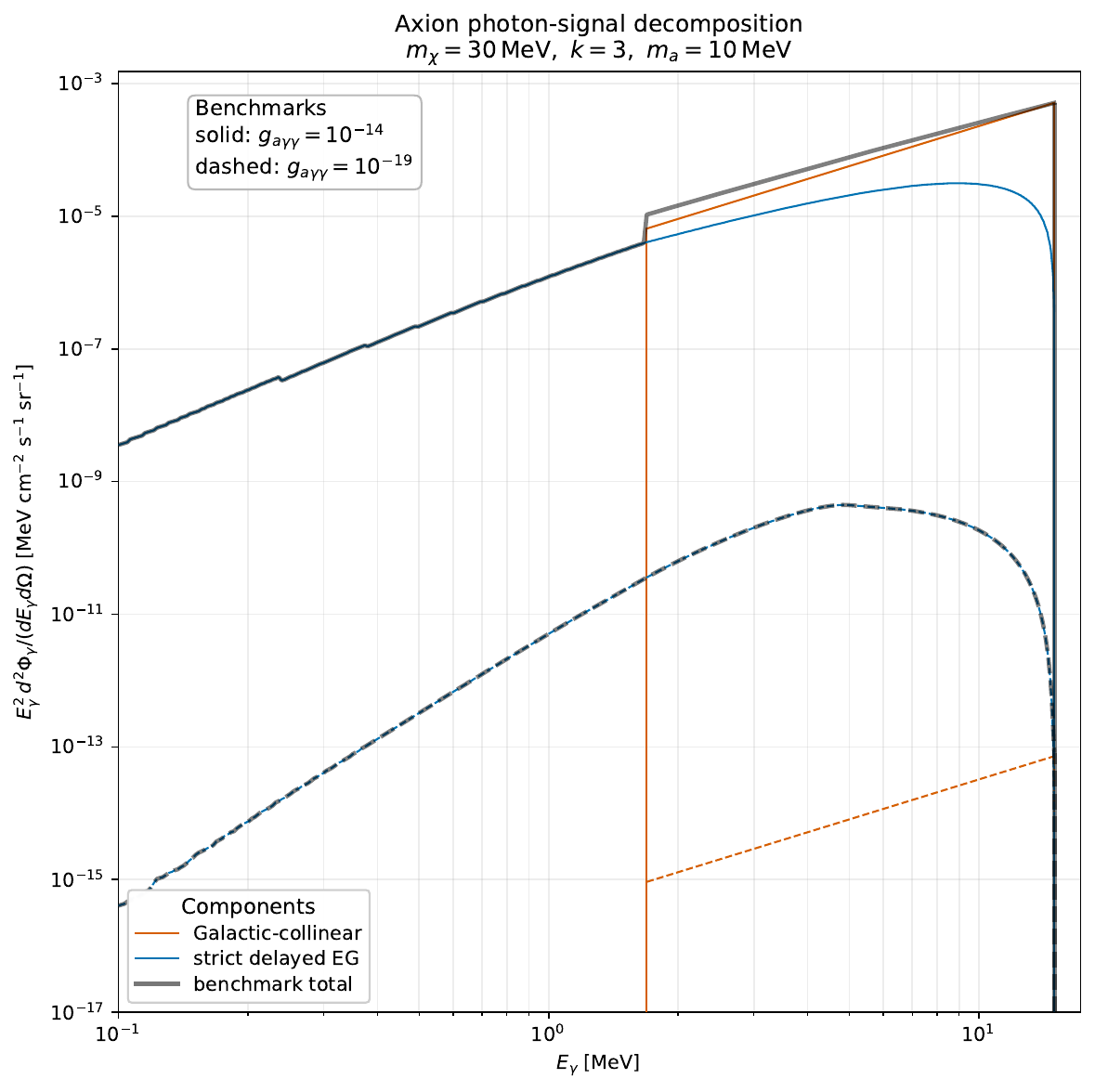}
\includegraphics[width=0.49\textwidth]{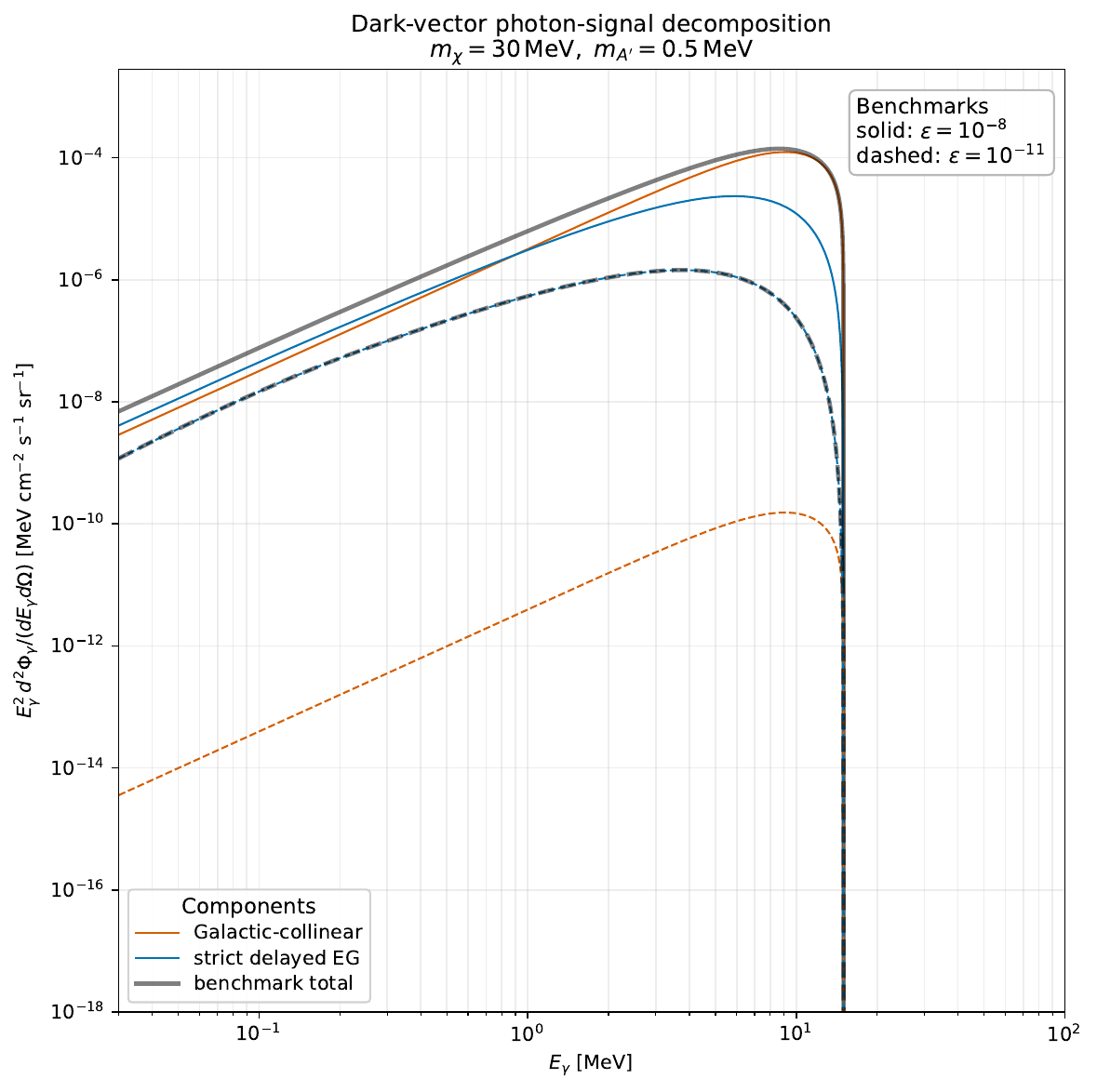}
\caption{Decomposition of the benchmark photon SED into Galactic-collinear and strict delayed EG components, All SEDs are normalized to $\tau_\chi=1.0\times10^{28}~{\rm s}$. Left: axion-like mediator benchmark with $m_\chi=30~{\rm MeV}$, $k=3$, and $m_a=10~{\rm MeV}$. Right: dark-vector mediator benchmark with $m_\chi=30~{\rm MeV}$ and $m_{A'}=0.5~{\rm MeV}$. In the long-lived regime, the EG component can become comparable to or larger than the Galactic-collinear contribution.}
\label{fig:sed_decomposition}
\end{figure}

In addition to the Galactic photon contribution, DM decays at cosmological distances give rise to an isotropic EG photon flux. For a long-lived mediator, the EG signal is intrinsically delayed: the DM particle may decay at redshift $z_p$, producing the mediator $X$, which subsequently decays at a lower redshift $z_d\leq z_p$. Within a homogeneous EG treatment, and assuming that the mediator momentum redshifts only due to cosmic expansion, we include the following delayed EG component in our benchmark photon flux:
\begin{align}
\frac{d^2\Phi_\gamma^{\rm EG,strict}}{dE_0\,d\Omega}=&\,\frac{1}{4\pi}\frac{\Omega_\chi\rho_c}{m_\chi\tau_\chi}\int_0^{z_{\rm max}}\frac{dz_d}{H(z_d)}\int_{z_d}^{z_{\rm max}}\frac{dz_p}{(1+z_p)H(z_p)}\nonumber\\
&\times\frac{\Gamma_X}{\gamma_X(z_d;z_p)}S_X(z_d,z_p)\,\frac{dN_\gamma}{dE_\gamma^{\rm em}}\left(E_\gamma^{\rm em};\gamma_X(z_d;z_p)\right)\,e^{-\tau_{\gamma\gamma}(E_0,z_d)} .
\label{eq:eg_cascade_general}
\end{align}
using the standard cosmological parameters of Ref.~\cite{Arguelles:2019ouk}, and we truncate the production-redshift integral at $z_{\rm max}=100$. Here $E_0$ is the observed photon energy, while $E_\gamma^{\rm em}=(1+z_d)E_0$ is the photon energy at the mediator decay epoch. The factor $S_X(z_d,z_p)$ denotes the probability that a mediator produced at $z_p$ survives until $z_d$, while $\Gamma_X/\gamma_X(z_d;z_p)$ is its time-dilated decay rate at the decay redshift, and $\gamma_X(z_d;z_p)$ is the mediator boost at the decay redshift after cosmological redshifting of its momentum. The exponential factor $e^{-\tau_{\gamma\gamma}(E_0,z_d)}$ accounts for photon attenuation during propagation, which we set to unity. A derivation of Eq.~\eqref{eq:eg_cascade_general}, together with the explicit definitions of $S_X$ and $\gamma_X(z_d;z_p)$, is given in Appendix~\ref{app:eg_cascade}. In the prompt-decay limit, this expression reduces to the standard EG decay flux, whereas in the cosmologically long-lived limit the flux scales linearly with the mediator width, $\Phi_\gamma^{\rm EG}\propto \Gamma_X$. 
The benchmark photon flux used in our main analysis is defined as the sum of the Galactic-collinear contribution and the strict delayed EG contribution,
\begin{align}
    \frac{d^2\Phi_{\gamma}^{\rm{bench}}}{dE_0\,d\Omega}=\frac{d^2\Phi_{\gamma}^{\rm{Gal,col}}}{dE_0\,d\Omega} + \frac{d^2\Phi_{\gamma}^{\rm{EG,strict}}}{dE_0\,d\Omega}
\label{eq:benchmark_sed}
\end{align}
Fig.~\ref{fig:sed_decomposition} shows that the delayed EG component can become comparable to or exceed the Galactic-collinear component for sufficiently long-lived mediators, we therefore retain both contributions in the benchmark photon flux.

\begin{figure}[htbp]
\centering
\includegraphics[width=0.49\textwidth]{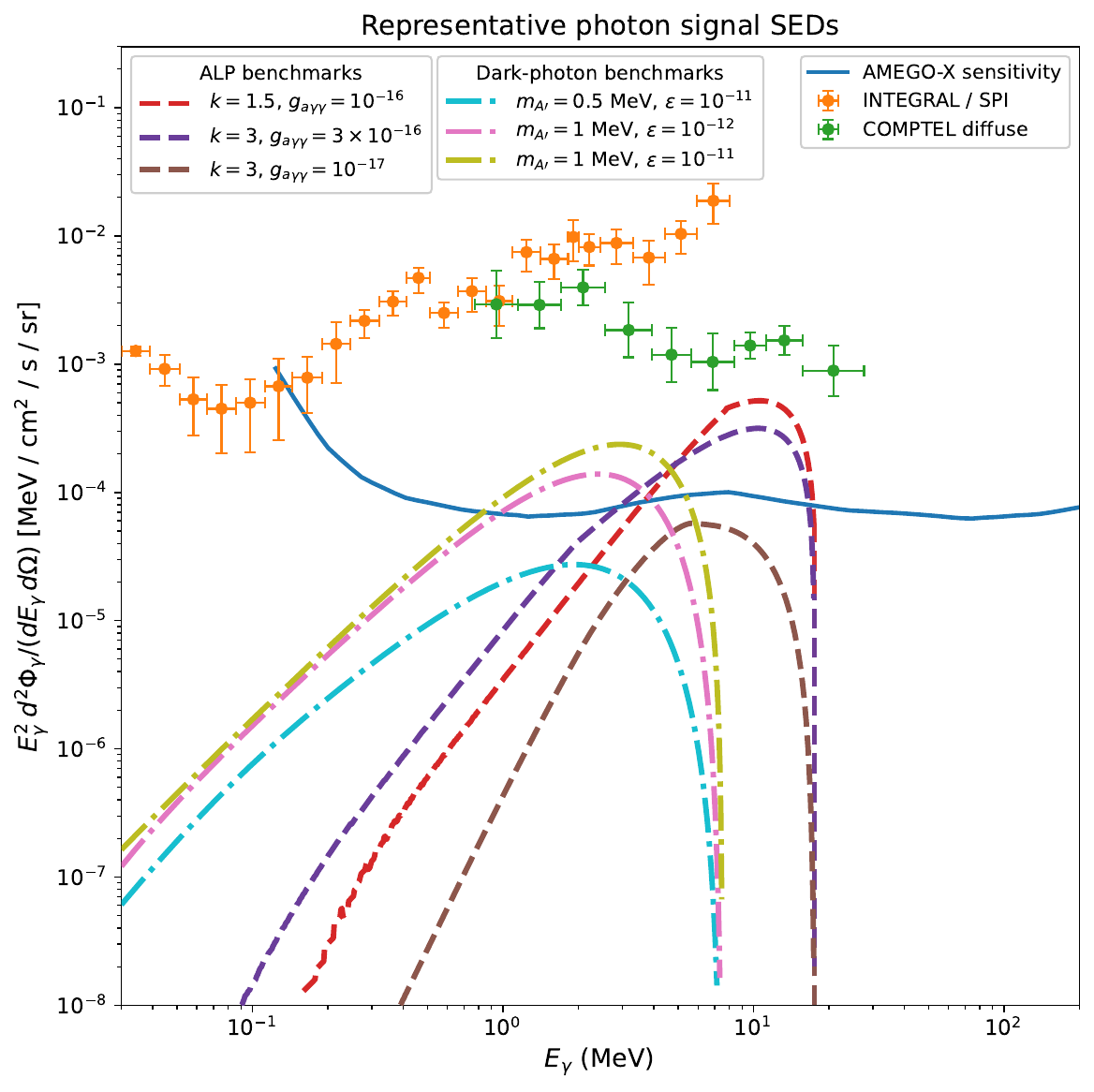}
\includegraphics[width=0.49\textwidth]{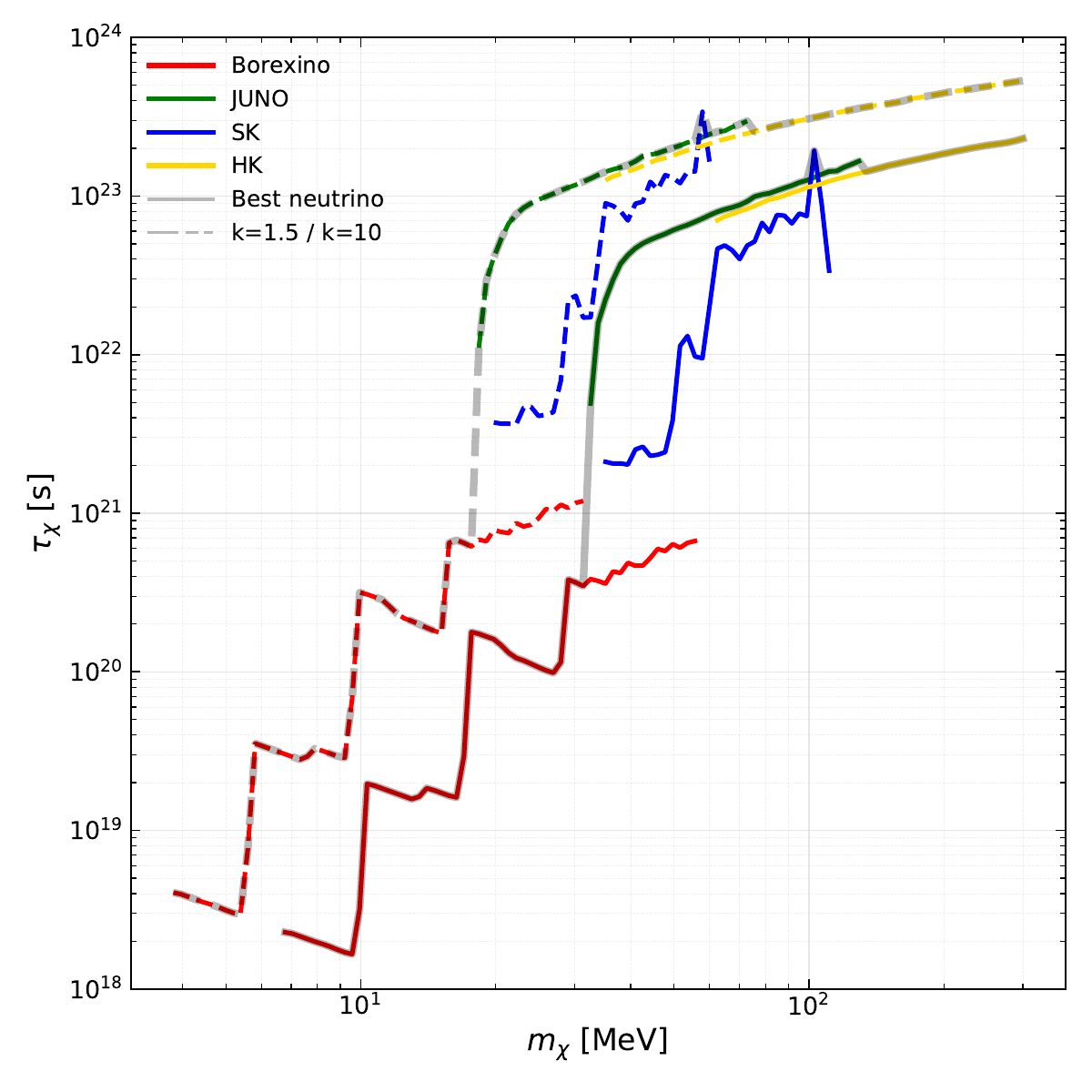}
\caption{Left: representative benchmark photon SEDs compared with INTEGRAL/SPI and COMPTEL data and the benchmark projected AMEGO-X reach. Theoretical spectra are shown for both axion-like particle $a \to \gamma\gamma$ ($m_\chi=35$ MeV) and dark-vector $A' \to 3\gamma$ ($m_\chi=15$ MeV) mediators. The displayed signals use the GC10 ROI for illustration and are normalized to $\tau_\chi=1.0\times10^{28}~{\rm s}$; the limits use the ROI of each data set. Right: current and projected 90\% C.L. neutrino lifetime limits for $k=1.5$ and $k=10$, showing the shift induced by the line-energy relation.}
\label{spectrum_neutrino}
\end{figure}

To derive 90\% confidence level (CL) lower limits on the DM lifetime $\tau_\chi = 1/\Gamma(\chi \to X + \nu)$, we employ two complementary approaches, depending on whether the observation is already available or projected for future instruments. For existing observations, we use:
\begin{itemize}
\item COMPTEL (Compton Telescope)~\cite{Essig:2013goa}, which operated in the 0.7--27\,MeV band. We use the reported diffuse flux measurements and uncertainties in the ROI $|b| < 20^\circ$ and $|l| < 60^\circ$.
\item INTEGRAL/SPI~\cite{Siegert:2022jii, Berteaud:2022tws}, which has accumulated more than 20 years of data in the 20\,keV--8\,MeV range. The ROI is taken to be $|b| < 47.5^\circ$ and $|l| < 47.5^\circ$.
\end{itemize}
For future missions, we choose:
\begin{itemize}
\item AMEGO-X (All-sky Medium Energy Gamma-ray Observatory eXplorer)~\cite{Caputo:2022xpx}, a proposed gamma-ray mission concept covering 100\,keV to 1\,GeV with significantly improved sensitivity.
\end{itemize}
To provide a more intuitive understanding of our recasting procedure, we compare the predicted DM-induced photon spectra directly with the observed diffuse gamma-ray data in the left panel of Fig.~\ref{spectrum_neutrino}, which shows representative observable photon SEDs after the $E_\gamma^2$ weighting and propagation effects are included. The box-shaped ALP injection spectrum and the smooth dark-vector three-body spectrum discussed in Sec.~\ref{sec:model} are therefore manifested mainly through different spectral endpoints, peak locations, and low-energy tails. Increasing the mass hierarchy broadens the energy distribution, whereas a longer mediator decay length suppresses the Galactic-collinear component, while the delayed EG contribution can remain important at intermediate lifetimes. As illustrated in Fig.~\ref{spectrum_neutrino}, current diffuse gamma-ray data from INTEGRAL and COMPTEL already exclude part of the parameter space, while future experiments such as AMEGO-X could substantially improve the sensitivity, extending the reach well beyond current limits in the MeV range.

For existing gamma-ray observations, we adopt a conservative background-agnostic recast, analogous to the method used for neutrino-flux limits in Ref.~\cite{Arguelles:2019ouk}. In this approach the measured diffuse flux in each bin is treated as an upper envelope for any additional dark-matter contribution, rather than as a background model to be subtracted. For the $i$-th energy bin we define a one-sided binned Gaussian statistic,
\begin{equation}
\chi^2_{\rm os}(\tau_\chi)
=
\sum_i
\left[
\frac{
\max\left(0,\Phi_i^{\rm DM}(\tau_\chi)-\Phi_i\right)
}{\sigma_i}
\right]^2 ,
\end{equation}
where $\Phi_i$ is the measured diffuse flux with uncertainty $\sigma_i$ and $\Phi_i^{\rm DM}$ is the predicted dark-matter contribution integrated over the same bin. This statistic penalizes only the part of the predicted signal that exceeds the observed flux, and therefore does not require an explicit model for the astrophysical background. The lower limit on the DM lifetime is obtained by requiring
\begin{equation}
\chi^2_{\rm os}(\tau_\chi)=n^2 .
\end{equation}
Throughout this work we quote one-sided Gaussian 90\% limits, corresponding to $n=1.28$.

For AMEGO-X, we follow the extended-source rescaling prescription of Ref.~\cite{Dutra:2025cwn}. Starting from the published three-year $3\sigma$ point-source continuum sensitivity of Ref.~\cite{Caputo:2022xpx}, we estimate the sensitivity to a $10^\circ$ disk around the Galactic Center as
\begin{equation}
E^2\frac{d\Phi_{\gamma,\rm sens}^{\rm ext}}{dE}=E^2\frac{d\Phi_{\gamma,\rm sens}^{\rm ps}}{dE}\left[\frac{\Delta\Omega_{10}}{\Delta\Omega_{\rm res}(E)}\right]^{1/4}
\end{equation}
where $E^2\frac{d\Phi_{\gamma,\rm sens}^{\rm ext}}{dE}$ ($E^2\frac{d\Phi_{\gamma,\rm sens}^{\rm ps}}{dE}$) denotes the estimated sensitivity for the extended target (point sources), $\Delta\Omega_{10}=2\pi(1-\cos10^\circ)\simeq 9.55\times10^{-2}\,{\rm sr}$ and $\Delta\Omega_{\rm res}(E)=\pi R_{\rm ARM/2}^2(E)$, where $R_{\rm ARM/2}(E)$ is the energy-dependent half width at half maximum of the angular-resolution-measure distribution. The corresponding GC10-averaged intensity sensitivity is
\begin{equation}
    E^2\frac{d^2\Phi_{\gamma,\rm sens}^{\rm ext}}{dE\,d\Omega} =\frac{1}{\Delta\Omega_{10}}E^2\frac{d\Phi_{\gamma,\rm sens}^{\rm ext}}{dE}
\end{equation}
We additionally rescale the published $3\sigma$ threshold by $1.28/3$ to obtain an approximate one-sided 90\% C.L. benchmark reach. 
For fixed dark-matter and mediator parameters, the predicted photon intensity scales inversely with the primary dark-matter lifetime. We therefore compute the benchmark photon SED at a reference lifetime $\tau_{\rm ref}$ and rescale it by a factor $\frac{\tau_{\rm ref}}{\tau_\chi}$. The benchmark AMEGO-X lifetime reach is obtained by requiring the predicted GC10-averaged SED not to exceed the rescaled sensitivity at any energy point. This prescription provides an approximate extended-source projection rather than a dedicated AMEGO-X likelihood for the dark-matter spatial template.

\section{Constraints from neutrino observations}
\label{sec:neutrino}
We also use the monochromatic neutrino line signal produced in the primary two-body decay to derive limits on the DM lifetime $\tau_\chi$, the neutrino is produced promptly and propagates essentially without attenuation. The intrinsic width of the line is set by the Galactic velocity dispersion, $\Delta E_\nu/E_\nu \sim v_{\rm DM} \sim 10^{-3}$, and is negligible compared with the experimental energy resolution. 
We therefore approximate the injected neutrino spectrum by the delta function in Eq.~\ref{eq:dndenu}, with the line energy given in Eq.~\ref{eq:E_nu}. We assume equal abundances of $\chi$ and $\bar{\chi}$, equal charge-conjugate decay rates, and equal contributions from the three light neutrino mass eigenstates to $\Gamma(\chi\to X\nu)$. By unitarity of the Pontecorvo--Maki--Nakagawa--Sakata (PMNS) matrix, this benchmark yields a flavor-democratic neutrino flux at Earth. The Galactic neutrino flux from DM decay is then
\begin{equation}
\frac{d\Phi_\nu}{dE_\nu}=\frac{f_\alpha}{4\pi m_\chi \tau_\chi}\frac{dN_\nu}{dE_\nu}D_{\rm DM},
\label{eq:nu_flux}
\end{equation}
We encode flavor effects through a factor $f_\alpha$, defined as the fraction of the total line flux constrained by a given experiment. The corresponding D-factor is given as
\begin{equation}
D_{\rm DM}=\int_{\Delta\Omega} d\Omega\int_{\rm l.o.s.} ds\,\rho_\chi(r(s,l,b))
\end{equation}
which is the standard decay $D$-factor. We use the same gNFW halo profile as in the photon analysis. The important difference from the photon case is that the neutrino flux is not multiplied by the mediator decay probability. 

We include the following neutrino inputs:
\begin{itemize}

\item 
\textbf{Borexino}: 
We use the model-independent upper limits on monochromatic electron-antineutrino 
fluxes from Ref.~\cite{Borexino:2019wln}. 
The relevant energy range is 
$E_\nu\simeq 1.8$--$16.8~\mathrm{MeV}$, corresponding to the inverse beta 
decay threshold and the energy window of the analysis. 
The constrained flavor is $\bar\nu_e$, and the detection channel is primarily
\begin{equation}
\bar\nu_e+p\to e^+ + n .
\end{equation}
We set $f_\alpha=1/6$ and since the experiment has no directionality, we apply it to the all-sky Galactic decay flux.

\item 
\textbf{JUNO}: 
For JUNO we use the projected sensitivity summarized in 
Ref.~\cite{Arguelles:2019ouk}. The energy range relevant for the projection is approximately 
$E_\nu\simeq 10$--$40~\mathrm{MeV}$, and the constrained channel is the electron-antineutrino component $\bar\nu_e$, detected mainly through inverse beta decay,
\begin{equation}
\bar\nu_e+p\to e^+ + n .
\end{equation}
In this window, reactor antineutrinos dominate the background below 
$\sim 11~\mathrm{MeV}$, while atmospheric-neutrino backgrounds become 
important at higher energies. 
We use this as a direction-insensitive projected flux limit and apply it to the all-sky Galactic decay flux and set $f_\alpha=1/6$.

\item 
\textbf{Super-Kamiokande}: 
We use the low-energy SK-$\bar\nu_e$ input summarized in Ref.~\cite{Arguelles:2019ouk}, based on the SK relic-supernova-neutrino search of Ref.~\cite{Linyan2018}. The signal is an electron-antineutrino flux, $\bar\nu_e$, detected primarily through inverse beta decay. The SK-IV neutron-tagging analysis uses 2778 live days in the prompt-energy window $11.3$--$31.3~\mathrm{MeV}$, with an additional 1886-day sample in the $9.3$--$11.3~\mathrm{MeV}$ window after the trigger upgrade. Together with the SK-I/II/III 2853-day sample used at higher energies, this provides a model-independent 90\% C.L. upper limit on the diffuse $\bar\nu_e$ flux in the low-energy DSNB window, approximately $E_\nu\simeq 10$--$30~\mathrm{MeV}$. Since the original analysis is a diffuse $\bar\nu_e$ flux search rather than a Galactic-Centre directional analysis, we apply it to the $\bar\nu_e$ component of the all-sky Galactic decay flux and set $f_\alpha=1/6$.

\item 
\textbf{Hyper-Kamiokande}: 
For HK we use the five-year-equivalent projected sensitivity labeled ``HK (Bell et al.)'' in Ref.~\cite{Arguelles:2019ouk}, based on the dedicated 20-year Hyper-Kamiokande simulation of Ref.~\cite{Bell:2020rkw}. In the low-mass region relevant for our analysis, the sensitivity is driven mainly by the fully contained electron-flavor sample. We therefore interpret this input as constraining the $\nu_e+\bar\nu_e$ component of the Galactic neutrino-line flux. Under the flavor-democratic benchmark adopted above and equal neutrino/antineutrino production, this corresponds to $f_\alpha=1/3$.
\end{itemize}
We do not include the EG neutrino component, which would appear as a redshift broadened continuum rather than a monochromatic line. Its omission makes the neutrino limits conservative for the line-based comparison considered here.

Although these experimental limits are ultimately obtained from neutrino--electron scattering or inverse beta decay event rates, we use the published line-flux limits directly. For each experiment, we construct $\Phi_{\nu_\alpha}^{\rm lim}(E_\nu)$ from the tabulated or digitized line-flux upper-limit curve and interpolate it in energy. This is sufficient for our purpose, since the neutrino limits enter only through the monochromatic line flux evaluated at $E_\nu^{\rm line}$. For a line-flux upper limit $\Phi_\nu^{\rm lim}(E_\nu)$, the corresponding lower bound on the DM lifetime is
\begin{equation}
\tau_\chi
\geq
\frac{f_\alpha D_{\rm DM}}
{4\pi m_\chi\,\Phi_{\nu_\alpha}^{\rm lim}(E_\nu^{\rm line})}.
\label{eq:tau_nu_limit}
\end{equation}

The behavior of the neutrino constraints is primarily governed by the decay kinematics and the flux scaling. As illustrated in the right panel of Fig.~\ref{spectrum_neutrino}, increasing $k$ from 1.5 to 10 shifts the entire exclusion envelope toward lower DM masses $m_\chi$. This kinematic shift occurs because a larger mass hierarchy requires a smaller $m_\chi$ to produce a neutrino within the optimal energy window of the detectors. Furthermore, the lifetime limits become notably more stringent as $m_\chi$ decreases; this is a direct consequence of the flux definition in Eq.~\ref{eq:tau_nu_limit}, where for a given experimental flux sensitivity, the bound on $\tau_\chi$ scales as $1/m_\chi$. Unlike the photon limit, the neutrino constraint is independent of $g_{a\gamma\gamma}$ and $\varepsilon$ because the line is produced at the primary vertex.

\section{Comparison and multi-Messenger synergy}
\label{sec:synergy}
Having derived photon and neutrino constraints separately in 
Secs.~\ref{sec:photon} and \ref{sec:neutrino}, we now compare the maximal reach of the two messenger channels. In the numerical analysis we take $m_\chi\in [1, 300]\,\rm{MeV}$, which covers the region where both the prompt neutrino line and the secondary photon continuum are probed by MeV-scale neutrino and gamma-ray searches in our work. This is because the neutrino energy is $E_\nu^{\rm line}=m_\chi(1-1/k^2)/2$, while the photon continuum typically extends up to $\mathcal O(m_\chi/2)$. Masses below this range are limited by neutrino thresholds, whereas higher masses move the photon signal toward gamma-ray searches outside the MeV-focused analysis adopted here.

\begin{figure}[htbp]
\centering
\includegraphics[width=0.49\textwidth]{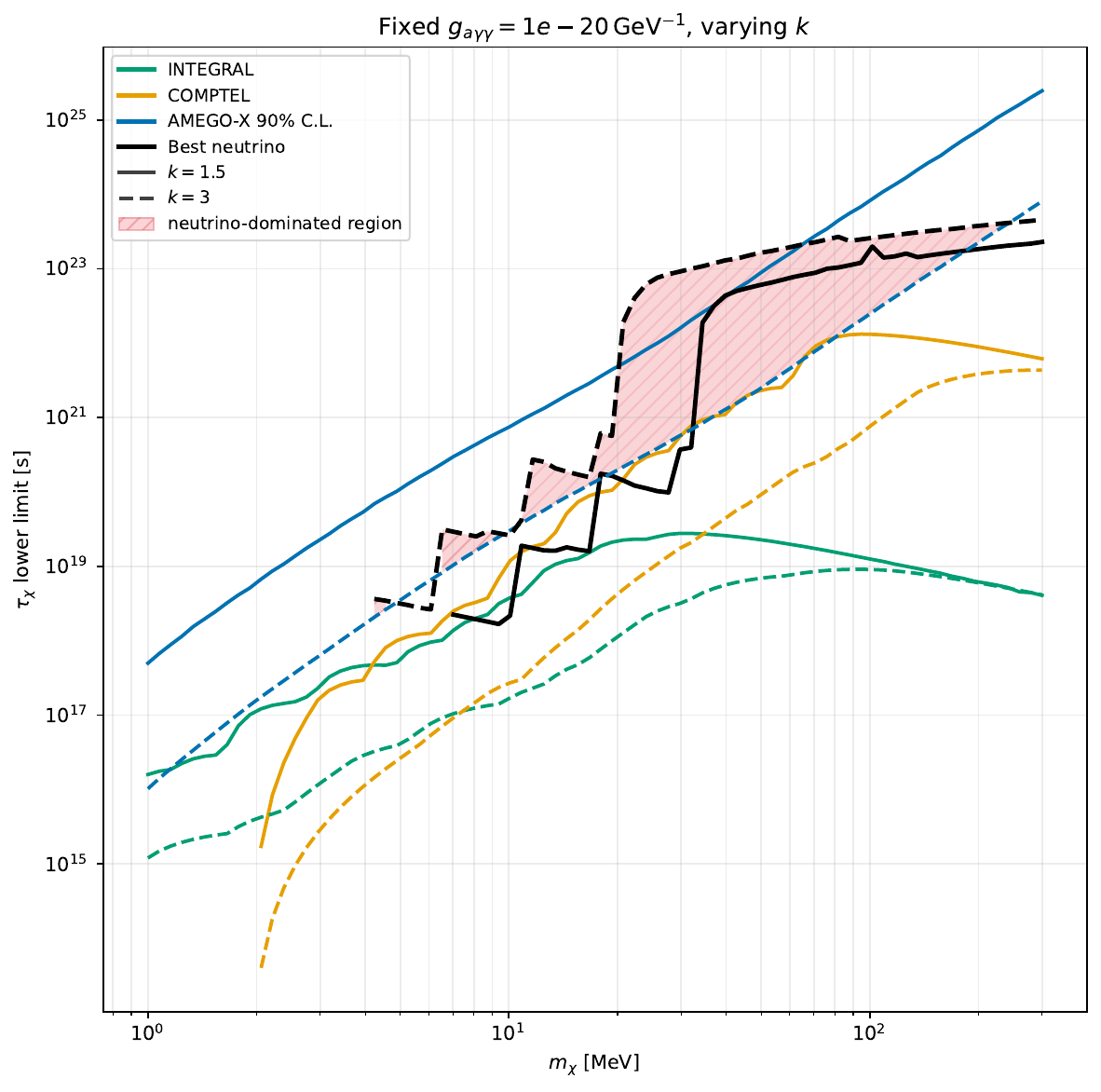}
\includegraphics[width=0.49\textwidth]{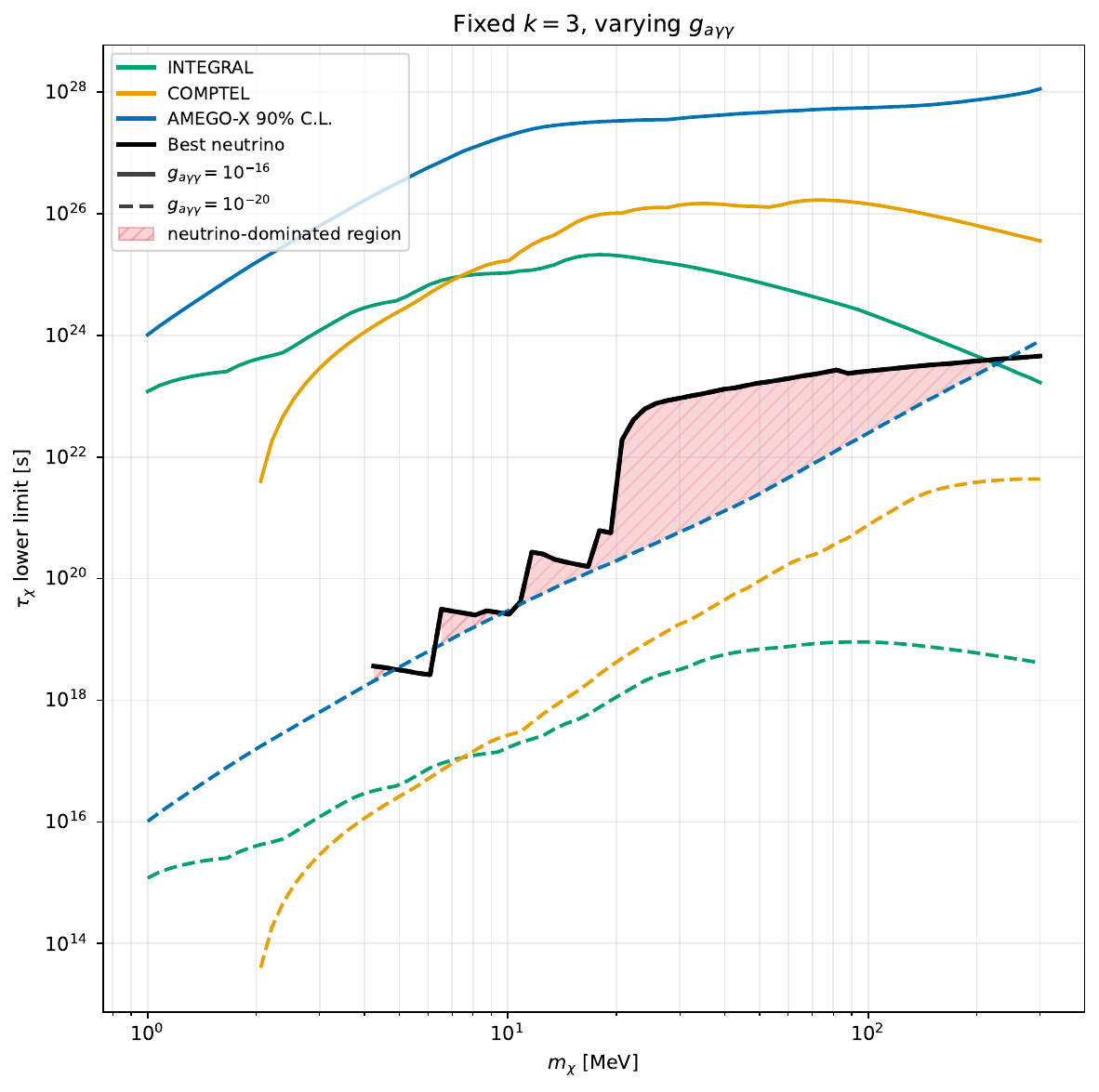}\\
\includegraphics[width=0.49\textwidth]{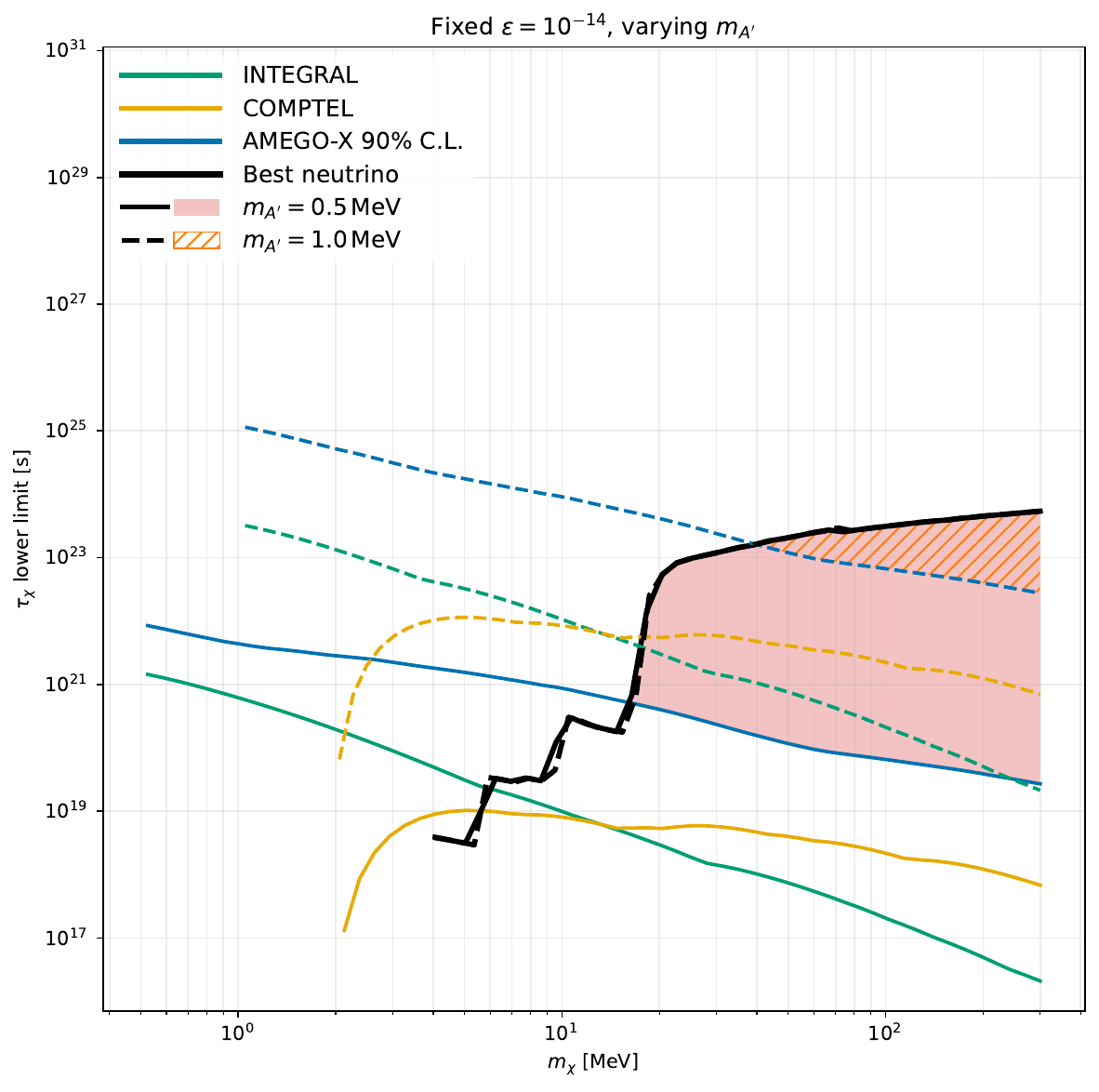}
\includegraphics[width=0.49\textwidth]{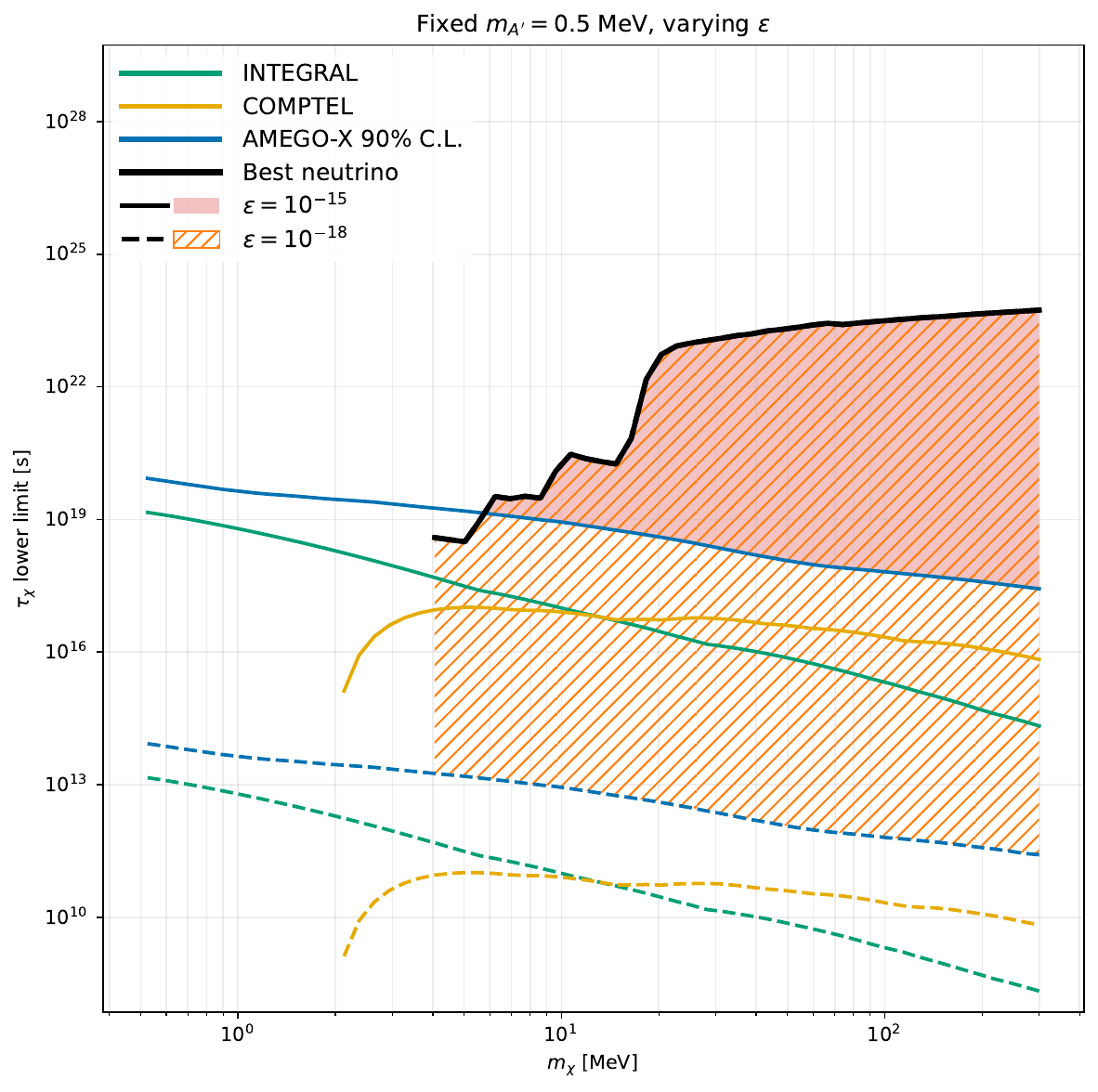}
\caption{
Comparison of photon and neutrino lifetime constraints. Upper panels: ALP mediator. Lower panels: dark-vector mediator. The neutrino curve corresponds to the best-neutrino envelope obtained from Borexino, Super-Kamiokande, JUNO, and Hyper-Kamiokande. Photon curves include current constraints from INTEGRAL/SPI and COMPTEL, together with the benchmark projected AMEGO-X reach. The shaded regions indicate where the best-neutrino envelope gives a stronger lifetime bound than all photon searches shown.
}
\label{fig:limit_comparison}
\end{figure}

We show the comparison for both the ALP and dark-vector mediator scenarios in Fig.~\ref{fig:limit_comparison}. In each panel, the shaded region denotes the parameter range where the best-neutrino envelope lies above all photon constraints, and therefore provides the leading bound on the DM lifetime. This construction allows us to identify where the neutrino line is not merely complementary, but becomes the dominant probe after the available photon searches are combined.

For the ALP mediator, the upper panels of Fig.~\ref{fig:limit_comparison} illustrate separately the dependence on the mass hierarchy and on the visible coupling. In the upper-left panel, $g_{a\gamma\gamma}$ is fixed while $k$ is increased from $1.5$ to $3$. Since $m_a=m_\chi/k$, a larger $k$ makes the ALP lighter and more boosted. The boosted $a\to\gamma\gamma$ box therefore becomes broader, redistributing the photon flux over a wider energy interval. At the same time, $\Gamma_{a\to\gamma\gamma}\propto g_{a\gamma\gamma}^2m_a^3$ decreases as $m_a$ becomes smaller, while the boost factor increases, so that the laboratory-frame decay length becomes longer. The neutrino envelope also shifts with $k$ for the purely kinematic reason discussed in Sec.~\ref{sec:neutrino}: a larger $k$ moves the same neutrino-energy window toward smaller $m_\chi$. The broader photon spectrum and longer mediator decay length weaken the photon constraints, while the kinematic shift of the neutrino envelope further changes the boundary of the neutrino-dominated region.

The upper-right panel keeps $k$ fixed and varies $g_{a\gamma\gamma}$. Decreasing the coupling increases the ALP lifetime and reduces the probability for the mediator to decay within the Galactic line-of-sight volume, thereby suppressing the Galactic-collinear photon component. The delayed EG contribution is also suppressed once the mediator becomes cosmologically long lived, although it can become comparable to or larger than the Galactic component before that limit is reached. The prompt neutrino line does not depend on $g_{a\gamma\gamma}$, so the neutrino-dominated region expands as the visible coupling is reduced.

The lower panels show the corresponding behavior for the dark-vector benchmark. In the lower-left panel, increasing the mediator mass from $m_{A'}=0.5~{\rm MeV}$ to $1.0~{\rm MeV}$ at fixed $\varepsilon$ has little effect on the neutrino envelope as long as $m_{A'}\ll m_\chi$, since then $E_\nu^{\rm line}\simeq m_\chi/2$. The photon signal is much more sensitive to this change: the heavier dark vector is less boosted and, more importantly, has a substantially larger decay width, $\Gamma_{A'\to3\gamma}\propto\varepsilon^2m_{A'}^9$ in the Euler--Heisenberg limit. Its decay is therefore more prompt, which strengthens the photon constraint and reduces the neutrino-dominated region. In the lower-right panel, decreasing $\varepsilon$ at fixed $m_{A'}$ suppresses the visible width and increases the propagation distance. The Galactic-collinear component is consequently reduced. The delayed EG contribution can remain important at intermediate lifetimes, but is also suppressed once the mediator becomes cosmologically long lived. The neutrino envelope is unchanged, so the neutrino-dominated region expands toward smaller $\varepsilon$.

\begin{figure}[htbp]
\centering
\includegraphics[width=0.49\textwidth]{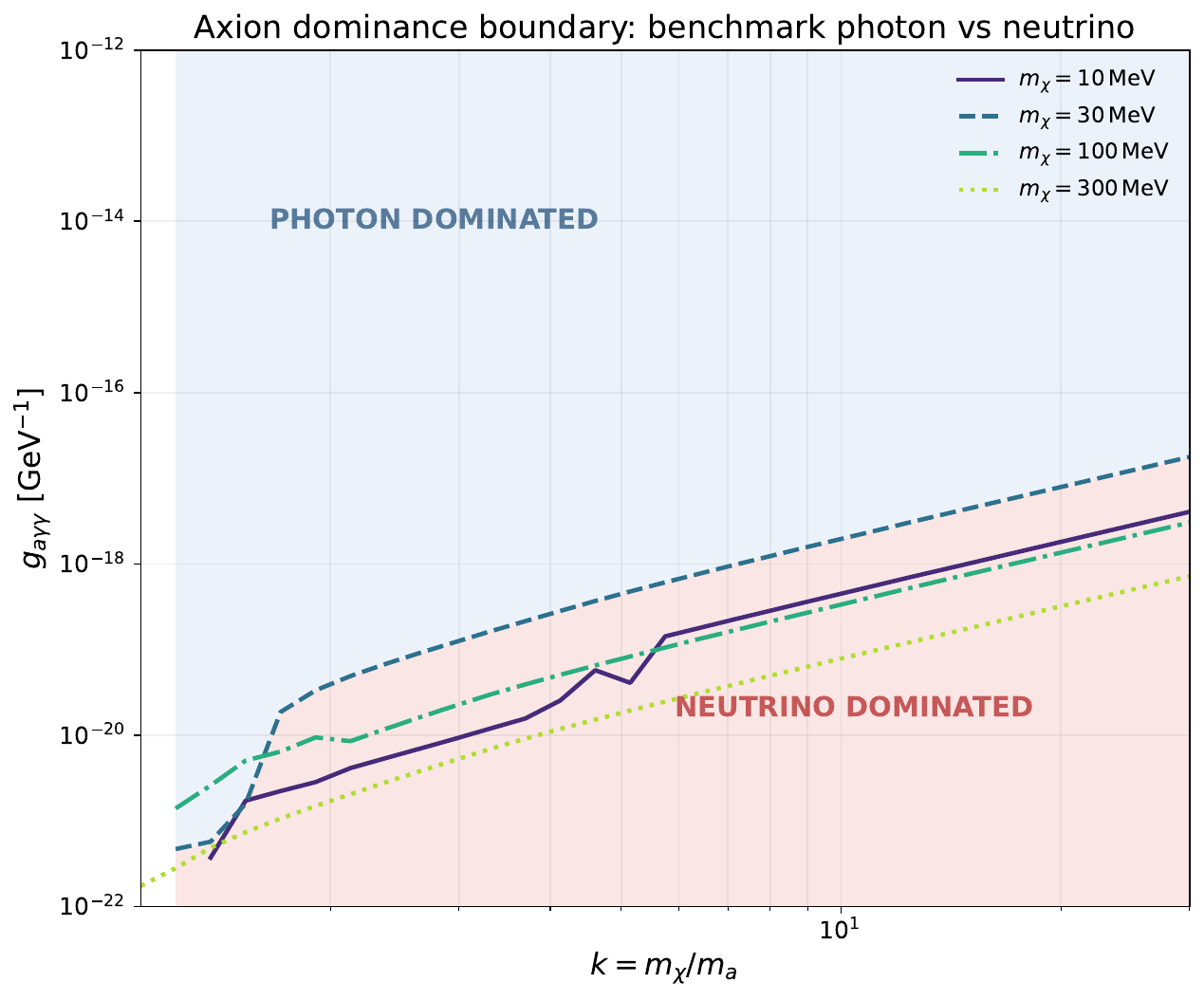}
\includegraphics[width=0.49\textwidth]{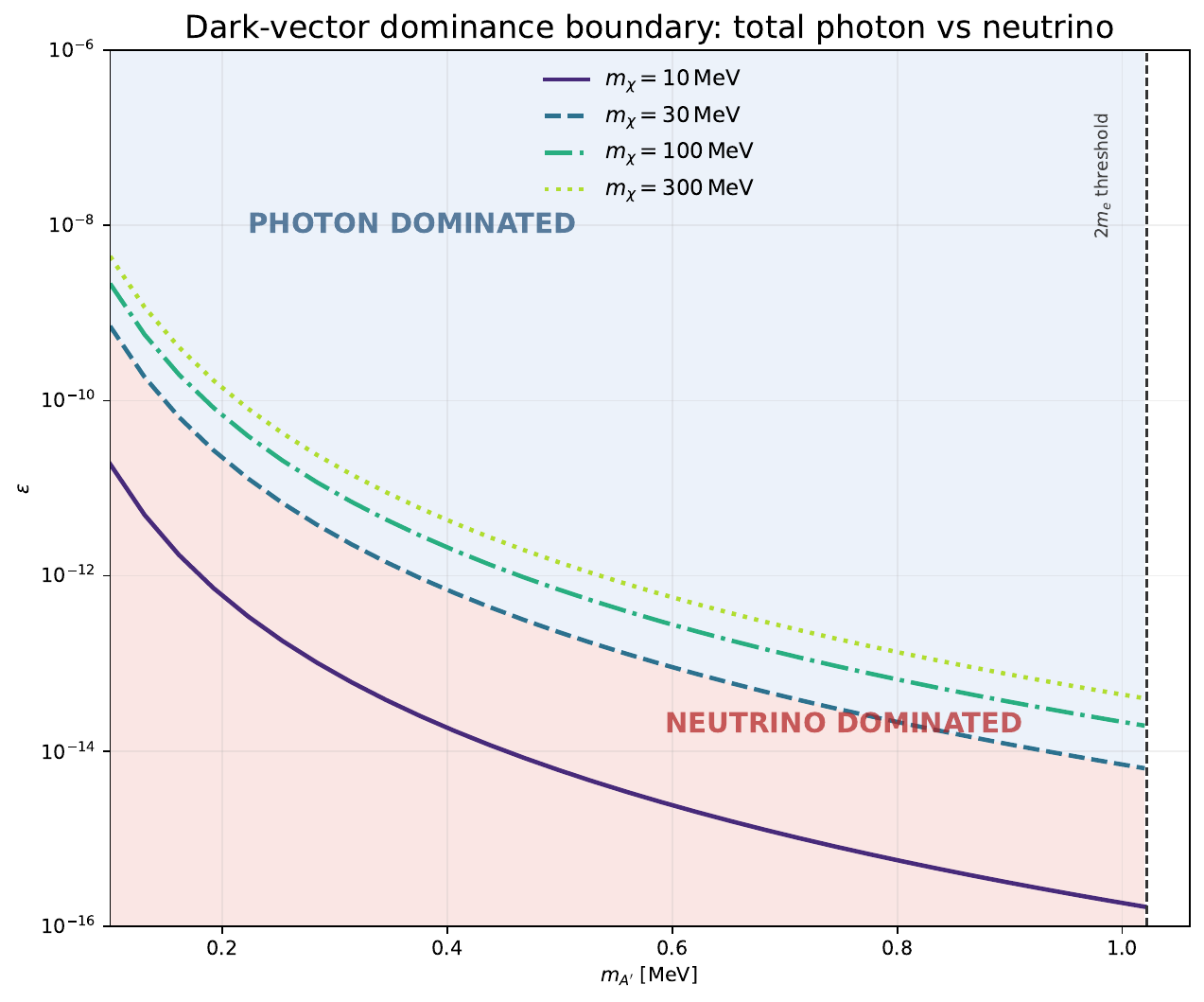}
\caption{
Dominance maps comparing the combined neutrino reach with the strongest photon constraints. Left: ALP mediator in the $(k,g_{a\gamma\gamma})$ plane, where $k=m_\chi/m_a$. Right: dark-vector mediator in the $(m_{A'},\varepsilon)$ plane, the vertical dotted line in the right panel marks the $A'=e^+e^-$ threshold, $m_{A'}=2m_e$. The contour lines correspond to the boundary $\tau_\nu^{\rm best}=\tau_\gamma^{\rm best}$ for representative DM masses $m_\chi=10,\,30,\,100,$ and $300~\mathrm{MeV}$. Here $\tau_\nu^{\rm best}$ denotes the best-neutrino envelope obtained from Borexino, Super-Kamiokande, JUNO, and Hyper-Kamiokande, while $\tau_\gamma^{\rm best}$ denotes the strongest photon constraint among INTEGRAL/SPI, COMPTEL, and AMEGO-X. The shaded regions indicate where the combined neutrino constraint is stronger than all photon constraints. In both mediator scenarios, the neutrino-dominated region appears in the long-lived regime, where the mediator decay length suppresses the observable photon flux inside the region of interest.
}
\label{fig:dominance_maps}
\end{figure}

In Fig.~\ref{fig:dominance_maps} we summarize the competition between the two messengers directly in the mediator parameter space. At each point, we compare the best-neutrino limit with the strongest benchmark photon limit, and the contours mark $\tau_\nu^{\rm best}=\tau_\gamma^{\rm bench,best}$. Over the parameter ranges shown, the photon envelope is typically dominated by the projected AMEGO-X reach. The shaded region therefore identifies where the prompt neutrino line gives a stronger lifetime constraint than all photon searches included in our analysis.

For the ALP benchmark, neutrino dominance is favored by larger $k$ and smaller $g_{a\gamma\gamma}$. At fixed $m_\chi$, increasing $k$ reduces $m_a=m_\chi/k$ and increases the mediator boost. Since $\Gamma_{a\to\gamma\gamma}\propto g_{a\gamma\gamma}^2m_a^3$, both effects increase the laboratory-frame decay length and suppress the Galactic photon signal. The broader boosted box spectrum also redistributes the photon flux over a wider energy range. Decreasing $g_{a\gamma\gamma}$ acts more directly by increasing the ALP lifetime. The neutrino line, in contrast, is unaffected by the visible coupling, so the dominance boundary moves toward larger regions of neutrino sensitivity as the mediator becomes longer lived.

The dark-vector case shows the same propagation-driven transition, but with a stronger dependence on the mediator mass. In the Euler--Heisenberg regime, $\Gamma_{A'\to3\gamma}\propto\varepsilon^2m_{A'}^9$, so decreasing either $m_{A'}$ or $\varepsilon$ rapidly increases the decay length and weakens the photon constraints. This produces the neutrino-dominated region toward small $m_{A'}$ and small $\varepsilon$. The different contour shapes in the ALP and dark-vector panels therefore reflect the different parameterization dependence of the visible widths, while the underlying transition is the same: increasing the mediator propagation length suppresses the Galactic photon component and eventually also the delayed EG contribution, whereas the primary neutrino line remains prompt.

We have also checked that replacing the benchmark photon flux by the strict-EG-only conservative flux does not change the dominance classification on the scanned benchmark grid within the recast assumptions adopted here. The corresponding diagnostic maps are shown in Appendix~\ref{app:EG}. This indicates that the appearance of the neutrino-dominated region is not tied to the Galactic-collinear benchmark, but is mainly controlled by the mediator propagation scale.

\section{Conclusion}
\label{sec:conclusion}

We have compared photon and neutrino constraints on sub-GeV dark matter undergoing the cascade decay $\chi\to X+\nu$, with $X=a$ or $A'$. By combining current and projected sensitivities from neutrino detectors (Borexino, Super-Kamiokande, JUNO, and Hyper-Kamiokande) with gamma-ray observations (COMPTEL, INTEGRAL/SPI, and AMEGO-X), we identified well-defined regions of parameter space in which neutrino-line searches provide the leading constraints on the DM lifetime.

The mediator propagation controls the relative reach of the two messengers. When the mediator decays promptly, the secondary photons approximately trace the DM distribution and gamma-ray observations provide strong constraints. As the mediator decay length increases, the Galactic-collinear component is suppressed, while the delayed EG contribution can remain important at intermediate lifetimes. The EG component is also suppressed once the mediator becomes cosmologically long lived. The prompt neutrino line remains unaffected by these propagation effects. This generates a robust mismatch between the photon and neutrino channels, even though both originate from the same primary decay process. The EG-only and spatial-smearing diagnostics indicate that this conclusion is stable on the scanned grid within the approximations of our analysis.

The quantitative transition depends on the mediator properties. The ALP case produces a boosted box-like spectrum from $a\to\gamma\gamma$, whereas the dark-vector case gives a softer three-body continuum from $A'\to3\gamma$. The corresponding visible widths also scale differently, with $\Gamma_{a\to\gamma\gamma}\propto g_{a\gamma\gamma}^2m_a^3$ and $\Gamma_{A'\to3\gamma}\propto\varepsilon^2m_{A'}^9$ up to loop corrections. These differences shift the quantitative location of the dominance boundary, but the underlying photon-to-neutrino transition is common to both scenarios. Future neutrino detectors, especially JUNO and Hyper-Kamiokande, can therefore provide competitive or leading sensitivity to cascade DDM in the long-lived mediator regime. Their combination therefore provides a useful test of sub-GeV cascade decays with long-lived mediators.

\begin{acknowledgments}
This work was supported by the National Natural Science Foundation of China under Grant No. 12305111, 12275232,  by the Jiangxi Provincial Natural Science Foundation 20252BAC200168,  by the Jiangxi Provincial Department of Education Scientific Research Program No. GJJ2200375, and by the National Gravitation Laboratory Open Research Projects NGL-2026-027.
\end{acknowledgments}
\appendix
\section{EG cascade signal from a long-lived mediator}
\label{app:eg_cascade}

In the main text we include the delayed EG component as part of the benchmark photon flux. Here we summarize its derivation and limiting behaviors. We work in a homogeneous EG approximation, neglect the depletion of the parent DM abundance, and assume that the mediator propagates freely until it decays. The latter assumption implies that the physical momentum of the mediator redshifts only due to cosmic expansion.
This component differs from the standard EG flux from prompt DM decay, because in the present scenario the photons are not produced at the primary DM decay redshift. Instead, the relevant process is
\begin{equation}
\chi(z_p)\to X(z_p)+\nu,
\qquad
X(z_p)\ {\rm propagates\ to}\ X(z_d),
\qquad z_d<z_p ,
\end{equation}
where $z_p$ denotes the production redshift of the mediator $X$, while
$z_d$ denotes its decay redshift. 
Consequently, the photon emission redshift is $z_d$, not $z_p$. 
A self-consistent EG treatment therefore requires a convolution over
both $z_p$ and $z_d$. In the rest frame of the DM particle, the mediator energy and momentum
at production are
\begin{equation}
E_{X,p}
=
\frac{m_\chi^2+m_X^2}{2m_\chi},
\qquad
p_{X,p}
=
\frac{m_\chi^2-m_X^2}{2m_\chi}.
\end{equation}
Equivalently, for $k\equiv m_\chi/m_X$,
\begin{equation}
\gamma_{X,p}
=
\frac{E_{X,p}}{m_X}
=
\frac{1}{2}\left(k+\frac{1}{k}\right),
\qquad
\beta_{X,p}\gamma_{X,p}
=
\frac{p_{X,p}}{m_X}
=
\frac{1}{2}\left(k-\frac{1}{k}\right).
\end{equation}
During cosmological propagation, the physical momentum of the mediator
redshifts as
\begin{equation}
p_X(z;z_p)
=
p_{X,p}\frac{1+z}{1+z_p}.
\end{equation}
The corresponding Lorentz factor at an intermediate redshift $z$ is therefore
\begin{equation}
\gamma_X(z;z_p)
=
\left[
1+
\frac{p_X^2(z;z_p)}{m_X^2}
\right]^{1/2}.
\label{eq:gamma_X_z}
\end{equation}

Let $\Gamma_X$ be the rest-frame decay width of the mediator. 
The decay rate in the cosmological frame is time-dilated by the factor
$\gamma_X(z;z_p)$. 
Using
\begin{equation}
dt
=
-\frac{dz}{(1+z)H(z)},
\end{equation}
the survival probability for a mediator produced at $z_p$ to remain
undecayed until $z_d$ is
\begin{equation}
S_X(z_d,z_p)
=
\exp\left[
-
\int_{z_d}^{z_p}
\frac{dz}{(1+z)H(z)}
\frac{\Gamma_X}{\gamma_X(z;z_p)}
\right].
\label{eq:survival_X}
\end{equation}
The probability density for the mediator to decay at $z_d$ is then
\begin{equation}
\frac{dP_X}{dt_d}(z_d|z_p)
=
\frac{\Gamma_X}{\gamma_X(z_d;z_p)}
S_X(z_d,z_p).
\label{eq:decay_pdf_X}
\end{equation}

The isotropic EG photon intensity can now be written as
\begin{align}
\frac{d^2\Phi_\gamma^{\rm EG}}{dE_0\,d\Omega}
=&\,
\frac{1}{4\pi}
\frac{\Omega_\chi\rho_c}{m_\chi\tau_\chi}
\int_0^{z_{\rm max}}
\frac{dz_d}{H(z_d)}
\int_{z_d}^{z_{\rm max}}
\frac{dz_p}{(1+z_p)H(z_p)}
\nonumber\\
&\times
\frac{\Gamma_X}{\gamma_X(z_d;z_p)}
S_X(z_d,z_p)\,
\frac{dN_\gamma}{dE_\gamma^{\rm em}}
\left(
E_\gamma^{\rm em};
\gamma_X(z_d;z_p)
\right)
\,e^{-\tau_{\gamma\gamma}(E_0,z_d)} .
\label{eq:eg_cascade_general_app}
\end{align}
Here $E_0$ is the photon energy observed today, while
\begin{equation}
E_\gamma^{\rm em}
=
(1+z_d)E_0
\end{equation}
is the photon energy at emission. 
The factor $e^{-\tau_{\gamma\gamma}}$ accounts for possible absorption during
propagation; in the MeV range considered in this work, this attenuation is
negligible and we set it to unity. 
We also assume $\tau_\chi\gg t_0$, so that the depletion of the parent dark
matter abundance over cosmic time can be neglected. 
If this condition is relaxed, the production rate in Eq.~\ref{eq:eg_cascade_general_app}
should be multiplied by the appropriate parent-density evolution factor.

Equation~\ref{eq:eg_cascade_general_app} has the correct prompt-decay limit.
When $\Gamma_X$ is large enough that the mediator decays immediately after
production, the decay probability density approaches
\begin{equation}
\frac{dP_X}{dt_d}(z_d|z_p)
\longrightarrow
\delta(t_d-t_p),
\end{equation}
or equivalently $z_d=z_p$. 
The double-redshift expression then reduces to the standard EG flux
for prompt DM decay,
\begin{equation}
\frac{d^2\Phi_\gamma^{\rm EG,prompt}}{dE_0\,d\Omega}
=
\frac{\Omega_\chi\rho_c}{4\pi m_\chi\tau_\chi}
\int_0^{z_{\rm max}} dz\,
\frac{1}{H(z)}
\frac{dN_\gamma}{dE_\gamma'}
\bigg|_{E_\gamma'=(1+z)E_0}.
\label{eq:prompt_limit}
\end{equation}
In the opposite long-lived limit, $S_X\simeq1$ and
$dP_X/dt_d\simeq \Gamma_X/\gamma_X$, so the EG cascade flux scales
linearly with $\Gamma_X$. 
Thus, once the decay length becomes larger than the Hubble scale, the isotropic
EG photon component is also suppressed. 
This behavior distinguishes the genuinely cosmologically long-lived regime
from the intermediate regime in which the mediator escapes the Galactic region
of interest but still decays on cosmological baselines.

\section{EG-only robustness of the dominance classification}
\label{app:EG}
In the main text we define the benchmark photon flux as Eq.~\ref{eq:benchmark_sed}. Since the Galactic part is treated in a collinear benchmark approximation, it is useful to test whether the photon/neutrino dominance classification depends on this component. 

We therefore compare the benchmark photon limit with a conservative EG-only photon limit, defined by
\begin{equation}
\frac{d^2\Phi_{\gamma}^{\rm EG-only}}{dE_0d\Omega}
\equiv
\frac{d^2\Phi_{\gamma}^{\rm EG,strict}}{dE_0d\Omega}.
\end{equation}
For comparison, we also define the Galactic-collinear-only photon limit from
\begin{equation}
\frac{d^2\Phi_{\gamma}^{\rm Gal-col-only}}{dE_0d\Omega}
\equiv
\frac{d^2\Phi_{\gamma}^{\rm Gal,col}}{dE_0d\Omega}.
\end{equation}
Therefore, we define the photon/neutrino dominance boundary obtained with the benchmark photon flux,
\begin{equation}
\tau_\nu^{\rm best}=\tau_\gamma^{\rm bench,best}.
\end{equation}
the conservative EG-only boundary,
\begin{equation}
\tau_\nu^{\rm best}=\tau_\gamma^{\rm EG-only,best}.
\end{equation}
and the EG-only photon limit equals the Galactic-collinear-only photon limit,
\begin{equation}
\tau_\gamma^{\rm EG-only,best}=\tau_\gamma^{\rm Gal-col-only,best}.
\end{equation}
All photon limits are obtained using the same experimental recast procedure as in the main text, including the corresponding energy grids and ROIs for INTEGRAL, COMPTEL, and AMEGO-X.
\begin{figure}[t]
\centering
\includegraphics[width=0.48\textwidth]{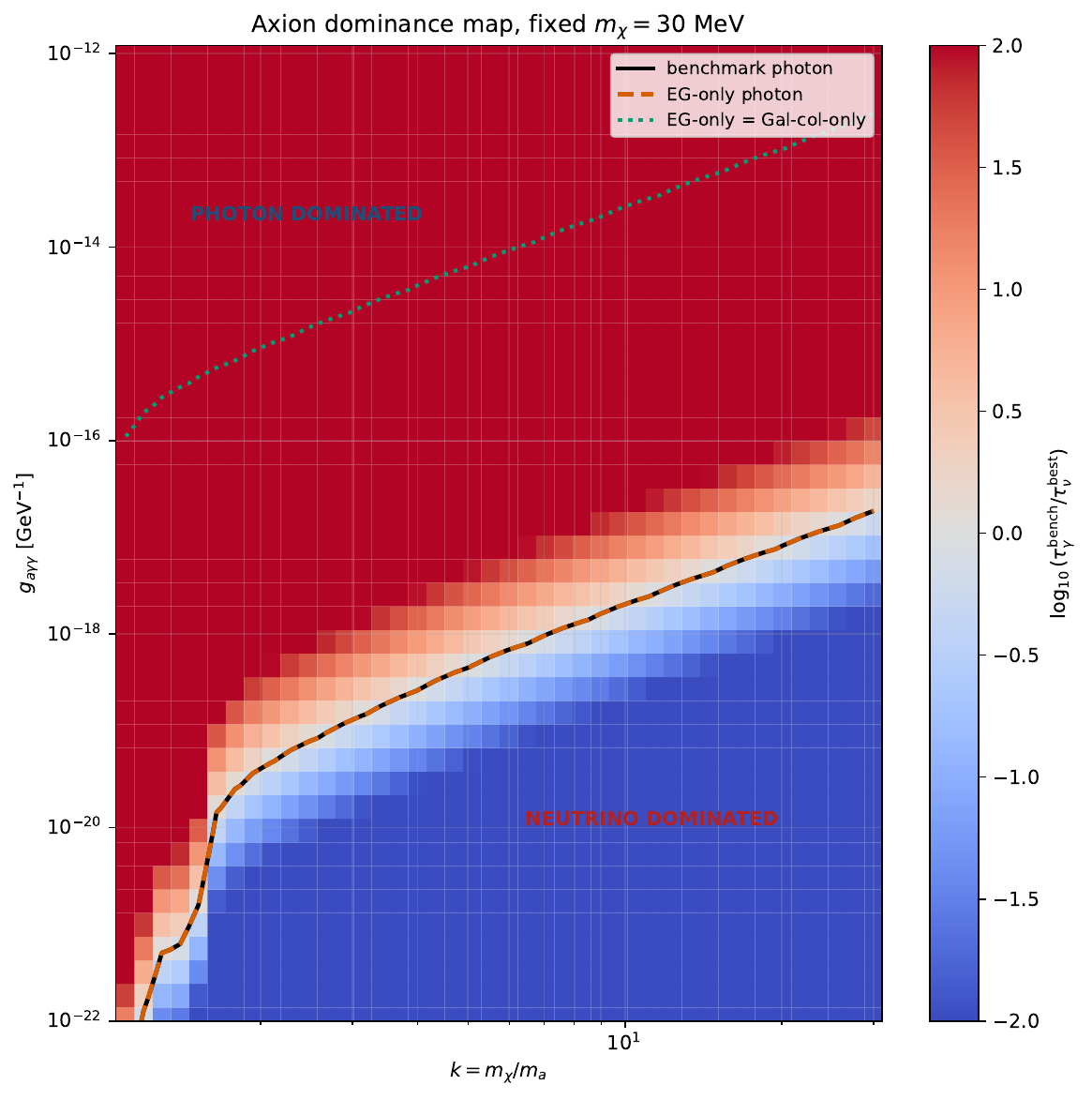}
\includegraphics[width=0.48\textwidth]{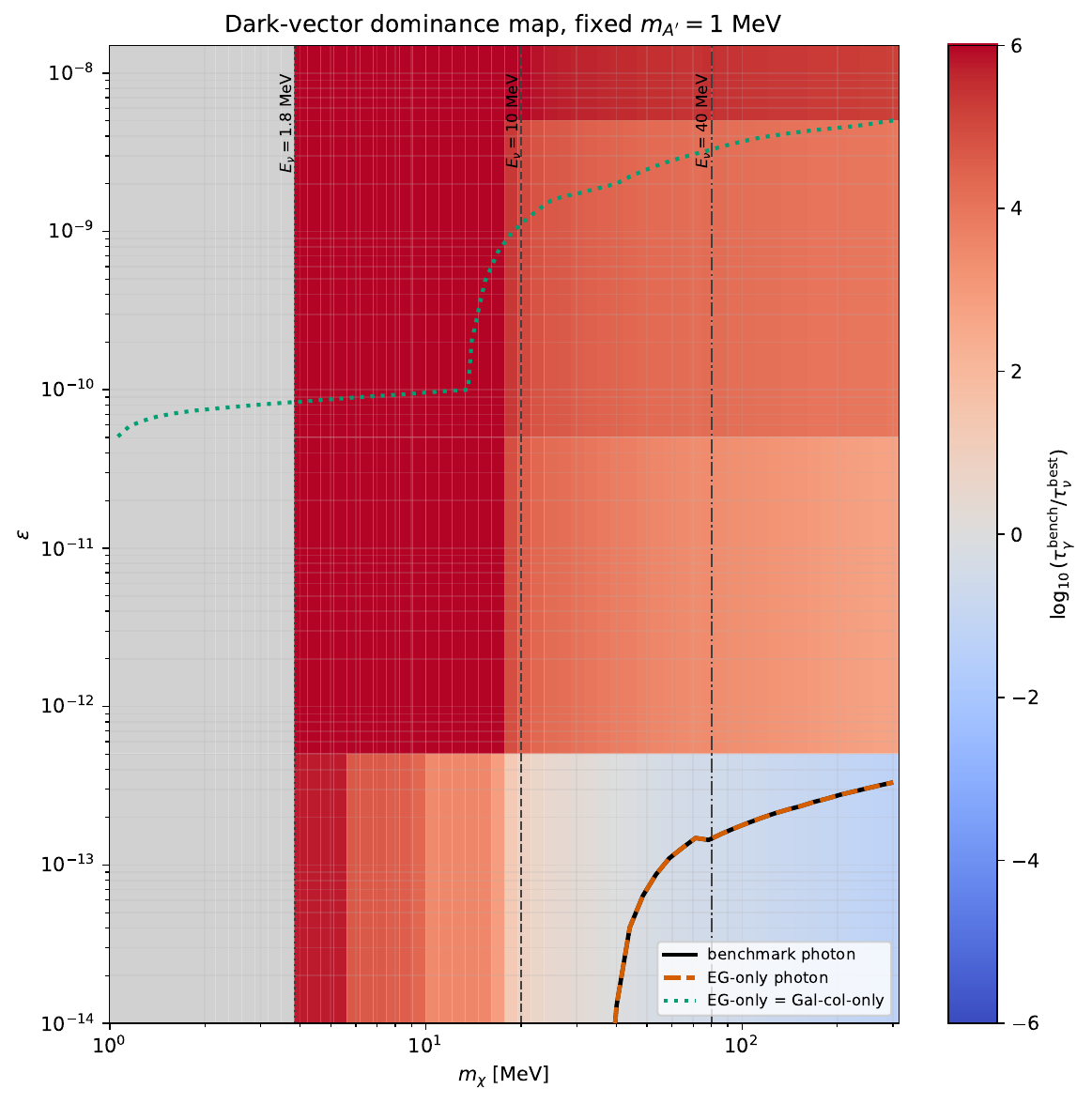}
\caption{EG-only robustness diagnostic for the photon/neutrino dominance classification. The background color shows $\log_{10}(\tau_\gamma^{\rm bench,best}/\tau_\nu^{\rm best})$, with positive (negative) values corresponding to photon (neutrino) dominance. The left panel fixes $m_\chi=30~{\rm MeV}$, while the right panel fixes $m_{A'}=1~{\rm MeV}$. The solid and dashed curves show the dominance boundaries obtained with the benchmark and EG-only photon fluxes, respectively, and the dotted curve marks $\tau_\gamma^{\rm EG-only,best} =\tau_\gamma^{\rm Gal-col-only,best}$. The vertical lines in the right panel correspond to $E_\nu^{\rm line}=1.8$, $10$, and $40~{\rm MeV}$.} 
\label{fig:robust}
\end{figure}
As shown in Fig.~\ref{fig:robust}, the benchmark and EG-only dominance boundaries nearly coincide over the parameter ranges considered. This indicates that the neutrino-dominated region is not an artifact of the Galactic-collinear benchmark within the recast assumptions adopted here.

\section{Galactic spatial-smearing validation}
\label{app:spatial_smearing}

The Galactic-collinear component used in the main analysis treats the mediator as propagating along the observed line of sight. This approximation captures the leading suppression from the finite decay probability, but it does not describe the full nonlocal redistribution of mediator decays in the Galactic halo. To test the Galactic-collinear approximation, we model the redistribution of mediator decay positions using the isotropic smearing kernel adapted from Ref.~\cite{Chu:2017vao},
\begin{equation}
\rho_{\rm smear}(\mathbf r_d)=\int d^3\mathbf r_p\rho_\chi(\mathbf r_p)\frac{\exp\left[-|\mathbf r_d-\mathbf r_p|/\lambda_X\right]}{4\pi \lambda_X |\mathbf r_d-\mathbf r_p|^2}.
\label{eq:rho_smear}
\end{equation}
The kernel includes the redistribution of decay positions but not the angular energy correlations of the daughter photons, we use it only to test the Galactic normalization. Here $\mathbf r_p$ is the mediator production point, $\mathbf r_d$ is the mediator decay point, and $\lambda_X$ is the boosted decay length. 

The nonlocal redistribution of Galactic mediator decays is controlled primarily by the laboratory-frame decay length $\lambda_X=\beta_X\gamma_X \tau_X$. For the ALP benchmark, with $m_a=m_\chi/k$,
\begin{equation}
\lambda_a\simeq17~{\rm kpc}\left(\frac{k^2(k^2-1)}{72}\right)\left(\frac{10^{-14}~{\rm GeV}^{-1}}{g_{a\gamma\gamma}}\right)^2\left(\frac{30~{\rm MeV}}{m_\chi}\right)^3 .
\end{equation}
While for the dark-vector case, in the Euler--Heisenberg and large-hierarchy limits,
\begin{equation}
\lambda_{A'}\simeq34~{\rm kpc}\left(\frac{m_\chi}{30~{\rm MeV}}\right)\left(\frac{10^{-8}}{\varepsilon}\right)^2\left(\frac{0.5~{\rm MeV}}{m_{A'}}\right)^{10} .
\end{equation}

The kernel in Eq.~\eqref{eq:rho_smear} is normalized so that, in an infinite volume, the total number of mediator decays is conserved. The corresponding smeared Galactic (D)-factor is
\begin{equation}
D_{\rm smear}(\Delta\Omega,\lambda_X)=
\int_{\Delta\Omega}d\Omega\int_{\rm l.o.s.} ds\rho_{\rm smear}(\mathbf r_d).
\label{eq:d_smear}
\end{equation}
We then define the spatial correction factor
\begin{equation}
C_{\rm spatial}(\Delta\Omega,\lambda_X)=
\frac{
D_{\rm smear}(\Delta\Omega,\lambda_X)
}{
D_{\rm Gal,col}(\Delta\Omega,\lambda_X)
},
\label{eq:c_spatial}
\end{equation}
where $D_{\rm Gal,col}$ is the Galactic-collinear effective $D$-factor used in the main benchmark calculation. The standard prompt $D$-factor is recovered only in the limit $\lambda_X\to0$.

We perform the validation for the ROIs used in the photon analysis and for decay lengths in the range
\begin{equation}
0.03~{\rm kpc}\leq\lambda_X\leq100~{\rm kpc}.
\end{equation}
\begin{figure}[t]
\centering
\includegraphics[width=0.8\textwidth]{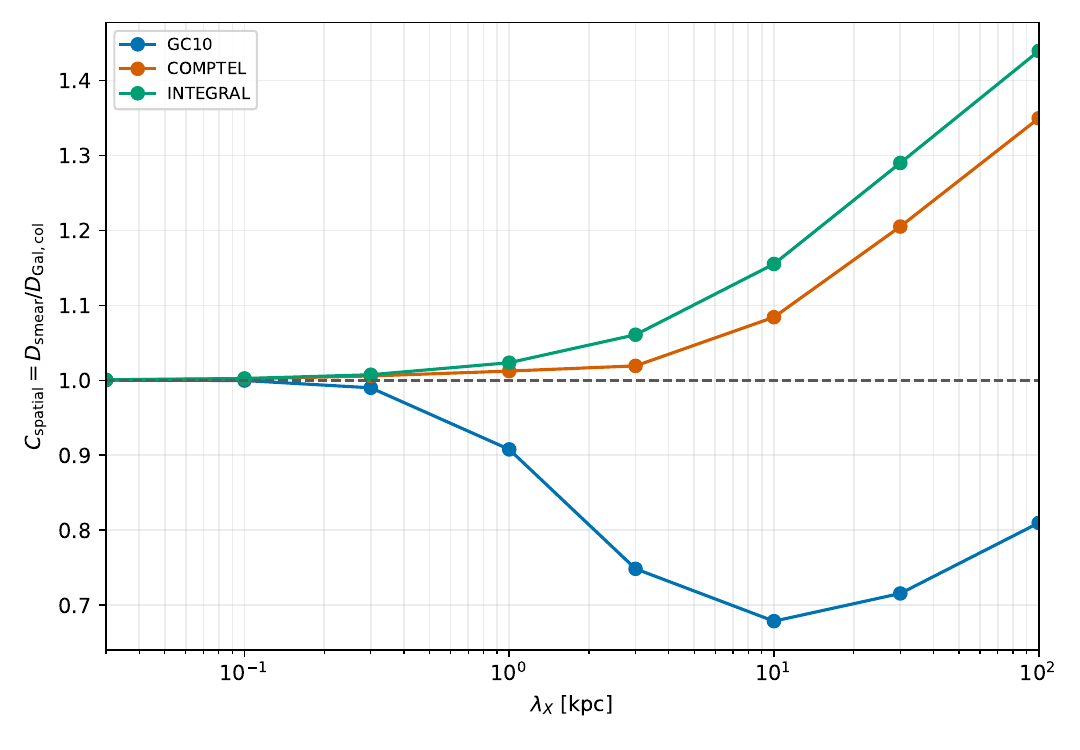}
\caption{Spatial-smearing correction factor $C_{\rm spatial}=D_{\rm smear}/D_{\rm Gal,col}$ as a function of $\lambda_X$ for the ROIs used in the photon analysis. The dashed horizontal line denotes
$C_{\rm spatial}=1$, corresponding to no correction relative to the Galactic-collinear benchmark. For Galactic-scale decay lengths, spatial smearing depletes compact Galactic-center regions such as GC10, while wider ROIs such as COMPTEL and INTEGRAL can receive an enhanced contribution from mediators produced outside the nominal line of sight and decaying inside the ROI. This correction is used only as a spatial validation of the Galactic normalization; it does not include the full angular--energy transport of the mediator decay products.
}
\label{fig:smear}
\end{figure}

We show the result in Fig.~\ref{fig:smear}. In the prompt limit, $C_{\rm spatial}\to1$, as expected. For Galactic-scale decay lengths, the correction can be order unity and depends on the ROI: compact Galactic-center regions are typically depleted by spatial smearing, while wider ROIs can receive contributions from mediators produced outside the nominal line of sight.

To assess the impact on the main result, we apply the spatial correction to the Galactic photon component and recompute the photon/neutrino dominance classification for all points whose decay lengths fall within the validated range. We find no change in the dominance classification among the conclusive points. This indicates that the neutrino-dominated regions identified in the main text are not an artifact of the Galactic-collinear benchmark normalization. Points with decay lengths beyond the validated range are instead covered by the EG-only conservative diagnostic discussed in Appendix~\ref{app:EG}.

\bibliography{main}

\end{document}